\documentclass[10pt, a4paper]{article}

\usepackage[english]{babel}
\usepackage{amsmath, latexsym}
\usepackage{amssymb}

\usepackage[authoryear, round]{natbib}
\usepackage{color}

\usepackage{ifpdf}

\ifpdf
   \usepackage{graphicx}
    \definecolor{myred}{rgb}{0.5,0,0}
    \definecolor{myblue}{rgb}{0,0,0.75}
    \definecolor{mygreen}{rgb}{0,0.5,0}
    \usepackage[pdftex,
               pdftitle={Capital Allocation to Business Units and Sub-Portfolios: the Euler Principle},
               pdfsubject={Risk measurement and management},
               pdfkeywords={credit portfolio model, capital requirement, quantile derivative, Euler's theorem, CDO modeling, risk impact},
               pdfauthor={Dirk Tasche},
               pdfstartview=FitR,
               bookmarks=false,
               breaklinks=true,
               colorlinks=true,
               citecolor=myred,
               linkcolor=myblue,
               urlcolor=mygreen]{hyperref}
\else
   \usepackage[dvips]{graphicx}
   \usepackage{rotating}
\fi

\newtheorem{theorem}{Theorem}[section]

\newtheorem{remark}{Remark}[section]

\newtheorem{proposition}{Proposition}[section]
\newtheorem{definition}{Definition}[section]
\newtheorem{corollary}{Corollary}[section]

\numberwithin{equation}{section}

\newlength{\captionwidth}
\begin{document}

\title{Capital Allocation to Business Units and Sub-Portfolios: the Euler Principle}

\author{%
Dirk Tasche\thanks{Lloyds TSB Bank, Corporate Markets. \newline
E-mail: dirk.tasche@gmx.net\newline
The opinions expressed in this chapter are those of the
author. They do not necessarily reflect views of Lloyds TSB Bank.}}

\date{\ }
\maketitle


\section{Introduction}

In many financial institutions, there is a well established practice
of measuring the risk of their portfolios in terms of
\emph{economic capital} \citep[cf., e.g.][]{Dev2004}.
Measuring portfolio-wide economic capital, however, is only the
first step towards active, portfolio-oriented risk management.
For purposes like identification of concentrations, risk-sensitive
pricing or portfolio optimization it is also necessary to decompose
portfolio-wide economic capital into a sum of risk contributions by
sub-portfolios or single exposures \citep[see, e.g.,][]{Litterman96}.

Overviews of a variety of different methodologies for this so-called \emph{capital allocation}
were given, e.g., by \citet{Koyluoglu&Stoker2002} and \citet{Urbanetal2004}.
\citet[][Section 6.3]{McNeil05} discuss in some detail the \emph{Euler allocation}
principle. The goal with this chapter is to provide more background information on the
Euler allocation, in particular with respect to the connection of Euler's theorem and
risk diversification and some estimation issues with Euler risk contributions. We furthermore investigate
a new application of Euler's theorem to the problem of how to identify the contributions of underlying names to
expected losses of collateralized debt obligation (CDO) tranches. \citet{MartinTasche} suggested to apply Euler allocation to measure the impact of systematic factors on portfolio risk. In the last section of the chapter, we illustrate this approach with another example and compare its results with the results of an alternative approach. The presentation of the subjects
is largely based on work on the subject by the author but, of course, refers to other
authors where appropriate.

Section \ref{sec:theory} ``Theory'' is mainly devoted to a motivation of the Euler allocation principle
by taking recourse to the economic concept of RORAC compatibility (Section \ref{se:contributions}).
Additionally, Section \ref{se:concentration} demonstrates that Euler risk contributions
are well-suited as a tool for the detection of risk concentrations. Section~\ref{sec:practice} ``Practice''
presents the formulae that are needed to calculate Euler risk contributions for standard deviation based risk
measures, Value-at-Risk (VaR) and Expected Shortfall (ES). Some VaR-specific estimation issues are
discussed in Section \ref{se:VaR}. Furthermore, in Section \ref{se:sample}, it is shown that there is a quite natural relationship between Euler contributions to VaR and the Nadaraya-Watson kernel estimator for conditional expectations. As a further practical application of
Euler's theorem, we analyze in Section \ref{se:CDO} the decomposition of CDO tranche expected loss into components associated with the underlying names and illustrate the results with a numerical example. In Section \ref{se:nl_impact} this example is re-used for illustrating different concepts of measuring the impact of single risk factors or sets of risk factors on portfolio risk. The chapter
concludes with a brief summary in Section \ref{sec:concl}.
Appendix \ref{sec:eulertheorem} provides some useful facts on homogeneous functions.

\section{Theory}
\label{sec:theory}

The Euler allocation principle may be applied to any risk measure that is homogeneous of
degree~1 in the sense of Definition \ref{de:homo} (see Appendix \ref{sec:eulertheorem}) and differentiable in an appropriate sense.
After having introduced the basic setting for the chapter in Section \ref{se:euler}, we discuss
in Section \ref{se:contributions} the economic motivation for the use of Euler risk contributions.
The fact that the Euler allocation principle can be derived by economic considerations constitutes the
-- maybe -- most appealing feature of this principle. We show in Sections \ref{se:subadd_contributions}
and \ref{se:concentration} that the economic foundation of the Euler allocation becomes even stronger in the case of sub-additive risk measures because it can then be used for efficient detection of risk concentration.

\subsection{Basic setting}
\label{se:euler}

Suppose that real-valued random variables $X_1, \ldots, X_n$ are given that
stand for the profits and losses with the assets (or some sub-portfolios) in a portfolio. Let $X$
denote the portfolio-wide profit/loss, i.e.\ let
\begin{equation}\label{eq:defY}
    X\ =\ \sum_{i=1}^n X_i.
\end{equation}
The economic capital ($\mathrm{EC}$) required by the portfolio (i.e.\ capital as
a buffer against high losses caused by the portfolio) is
determined with a \emph{risk measure} $\rho$, i.e.\
\begin{equation}\label{eq:ec}
    \mathrm{EC}\ =\ \rho(X).
\end{equation}
In practice, usually, $\rho$ is related to the variance or a quantile of the portfolio loss
distribution. See Section \ref{sec:practice} for some examples of how $\rho$ can be chosen.

For some kinds of risk (in particular credit risk), for risk management traditionally only losses
are considered. Let $L_i \ge 0$ denote the loss with (credit) asset $i$ and assume that $L =
\sum L_i$ stands for portfolio-wide loss. Then this loss perspective on portfolio risk can be
reconciled with the profit/loss perspective from \eqref{eq:defY} and \eqref{eq:ec} by
considering
\begin{equation}\label{eq:XL}
    X_i \ = \ g_i - L_i,
\end{equation}
where $g_i$ denotes the lender's stipulated gain with credit asset $i$ if the credit is repaid regularly.
In the following, as a general rule we will adopt the profit/loss perspective from \eqref{eq:ec}.

It is useful to allow for some dynamics in model \eqref{eq:defY} by introducing weight
variables $u = (u_1, \ldots, u_n)$:
\begin{equation}\label{eq:X(u)}
    X(u) \ = \ X(u_1, \ldots, u_n)\ =\ \sum_{i=1}^n u_i\,X_i.
\end{equation}
Then we have obviously $X = X(1, \ldots, 1)$. The variable $u_i$ can be interpreted as
amount of money invested in the asset which underlies $X_i$ or just as the credit exposure if
$X_i$ is chosen as a default indicator. For the purpose of this chapter we assume that the
probability distribution of the random vector $(X_1, \ldots, X_n)$ is fixed. We will, however,
consider some variations of the variable $u$. It is then convenient to introduce the function
\begin{equation}\label{eq:fct_risk}
    f_{\rho,X}(u)\ = \ \rho(X(u)).
\end{equation}
For the same risk measure $\rho$, the function $f_{\rho,X}$ can look quite different for different
distributions of $X$. As we assume that the distribution of $X$ is fixed, we can nevertheless drop
the index $X$ and write $f_\rho$ for $f_{\rho,X}$.

In this chapter, we focus attention to (positively) \emph{homogeneous} risk measures $\rho$ and
functions $f_\rho$ (see Appendix \ref{sec:eulertheorem} for some important properties of such functions).

\subsection{Defining risk contributions}
\label{se:contributions}

When portfolio-wide economic capital is measured according to \eqref{eq:ec}, it may be useful
to answer the question:
How much contributes asset (or sub-portfolio) $i$ to $\mathrm{EC} = \rho(X)$? Some potential
applications of the answer to this question will be given below. For the time being,
we denote the (still to be defined) \emph{risk contribution} of $X_i$ to $\rho(X)$ by $\rho(X_i \,|\,X)$.

\begin{definition}\label{de:rorac}
Let $\mu_i = \mathrm{E}[X_i]$. Then
\begin{itemize}
  \item the total portfolio \emph{\underbar{R}eturn \underbar{o}n \underbar{R}isk
    \underbar{A}djusted \underbar{C}apital}\footnote{%
Note that performance measurement by RORAC can be motivated by Markowitz-type risk-return optimization for general risk measures that are homogeneous of any degree $\tau$ \citep[see][Section 6]{Tasche1999}.%
}     is defined by
    \begin{equation*}
        \mathrm{RORAC}(X)\ =\ \frac{\mathrm{E}[X]}{\rho(X)}\ = \ \frac{\sum_{i=1}^m \mu_i}{\rho(X)},
    \end{equation*}
  \item the portfolio-related \emph{RORAC} of the $i$-th asset is defined by
  \begin{equation*}
    \mathrm{RORAC}(X_i\,|\,X)\ =\ \frac{\mathrm{E}[X_i]}{\rho(X_i \,|\,X)}\ =\ \frac{\mu_i}{\rho(X_i\,|\,X)}.
  \end{equation*}
\end{itemize}
\end{definition}

Based on the notion of RORAC as introduced in Definition \ref{de:rorac}, two properties
of risk contributions can be stated that are desirable from an economic point of view.

\begin{definition}\label{de:desirable}
Let $X$ denote portfolio-wide profit/loss as in \eqref{eq:defY}.
\begin{itemize}
  \item Risk contributions $\rho(X_1 \,|\,X), \ldots, \rho(X_n \,|\,X)$ to portfolio-wide risk
  $\rho(X)$ satisfy the \emph{full allocation} property if
  \begin{equation*}
    \sum_{i=1}^n \rho(X_i \,|\,X) \ = \ \rho(X).
  \end{equation*}
  \item Risk contributions $\rho(X_i \,|\,X)$ are \emph{RORAC compatible} if there are some $\epsilon_i > 0$ such that
  \begin{equation*}
\mathrm{RORAC}(X_i\,|\,X)\ >\ \mathrm{RORAC}(X) \qquad
\Rightarrow\qquad \mathrm{RORAC}(X+ h\,X_i) \ >\ \mathrm{RORAC}(X)
\end{equation*}
for all $0 < h < \epsilon_i$.
\end{itemize}
\end{definition}

It turns out that in the case of a ``smooth'' risk measure $\rho$, requiring the RORAC compatibility property
of Definition \ref{de:desirable} completely determines the risk contributions $\rho(X_i \,|\,X)$.

\begin{proposition}\label{pr:determ}
Let $\rho$ be a risk measure and $f_\rho$ be the function that corresponds to $\rho$ according to
\eqref{eq:X(u)} and \eqref{eq:fct_risk}. Assume that $f_\rho$ is continuously differentiable. If
there are risk contributions $\rho(X_1 \,|\,X), \ldots,$ $\rho(X_n \,|\,X)$ that are RORAC compatible
in the sense of Definition \ref{de:desirable} for arbitrary expected values $\mu_1, \ldots, \mu_n$ of
$X_1, \ldots, X_n$, then $\rho(X_i \,|\,X)$ is uniquely determined as
\begin{equation}\label{eq:deriv}
    \rho_{\mathrm{Euler}}(X_i\,|\,X)\ =\ \frac{d\, \rho}{d\,h}(X+h\,X_i)\bigl|_{h=0}\ =\
    \frac{\partial\, f_\rho}{\partial\,u_i}(1, \ldots, 1).
\end{equation}
\end{proposition}
See Theorem 4.4 of \citet{Tasche1999} for a proof of Proposition \ref{pr:determ}. It is easy to see that risk
contributions defined by \eqref{eq:deriv} are always RORAC compatible.

\setcounter{figure}{0}
\refstepcounter{figure}
\begin{figure}[tb]
  \begin{center}
  \parbox{14.0cm}{Figure \thefigure:
  \emph{Illustration of Definition \ref{de:desirable} and Proposition \ref{pr:determ}. Sub-portfolio profit/loss variables $X_1 = 0.015 - L_1$, 
  $X_2 = 0.04 - L_2$ with $L_1, L_2$ as defined by \eqref{eq:def_Li}. Portfolio-wide profit/loss $X = X(u) = u\,X_1 + (1-u)\,X_2$, where $u\in [0,1]$ is the weight of sub-portfolio~1. $\rho(X) = \mathrm{UL}_{\mathrm{VaR}, \alpha}(X)$ as defined by \eqref{eq:ULVaR}.
   Parameter values as in \eqref{eq:params}.}}
\label{fig:RORAC}\\[1ex]
\ifpdf
    \resizebox{\height}{11.0cm}{\includegraphics[width=11.0cm]{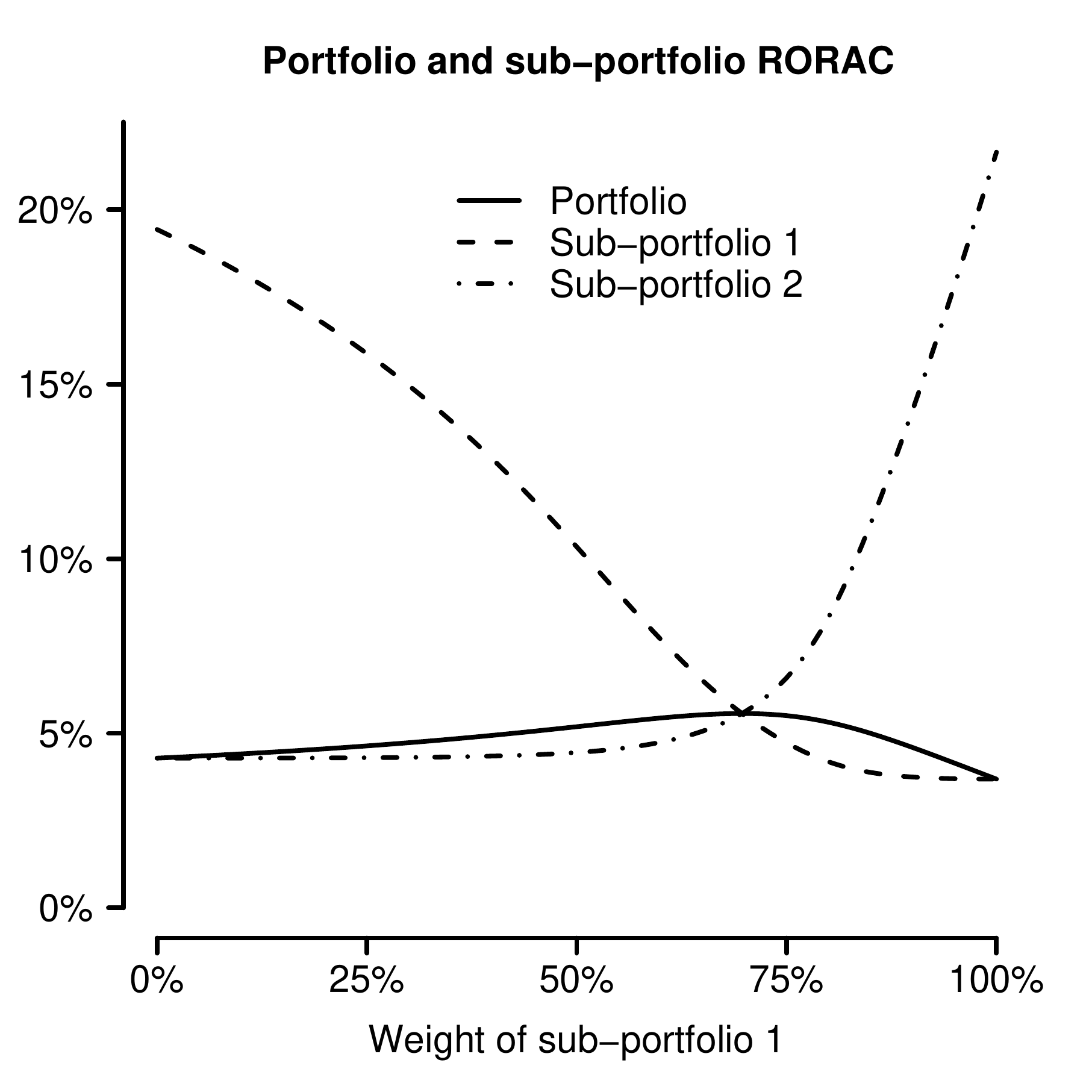}}
\else
\begin{turn}{270}
\resizebox{\height}{11.0cm}{\includegraphics[width=11.0cm]{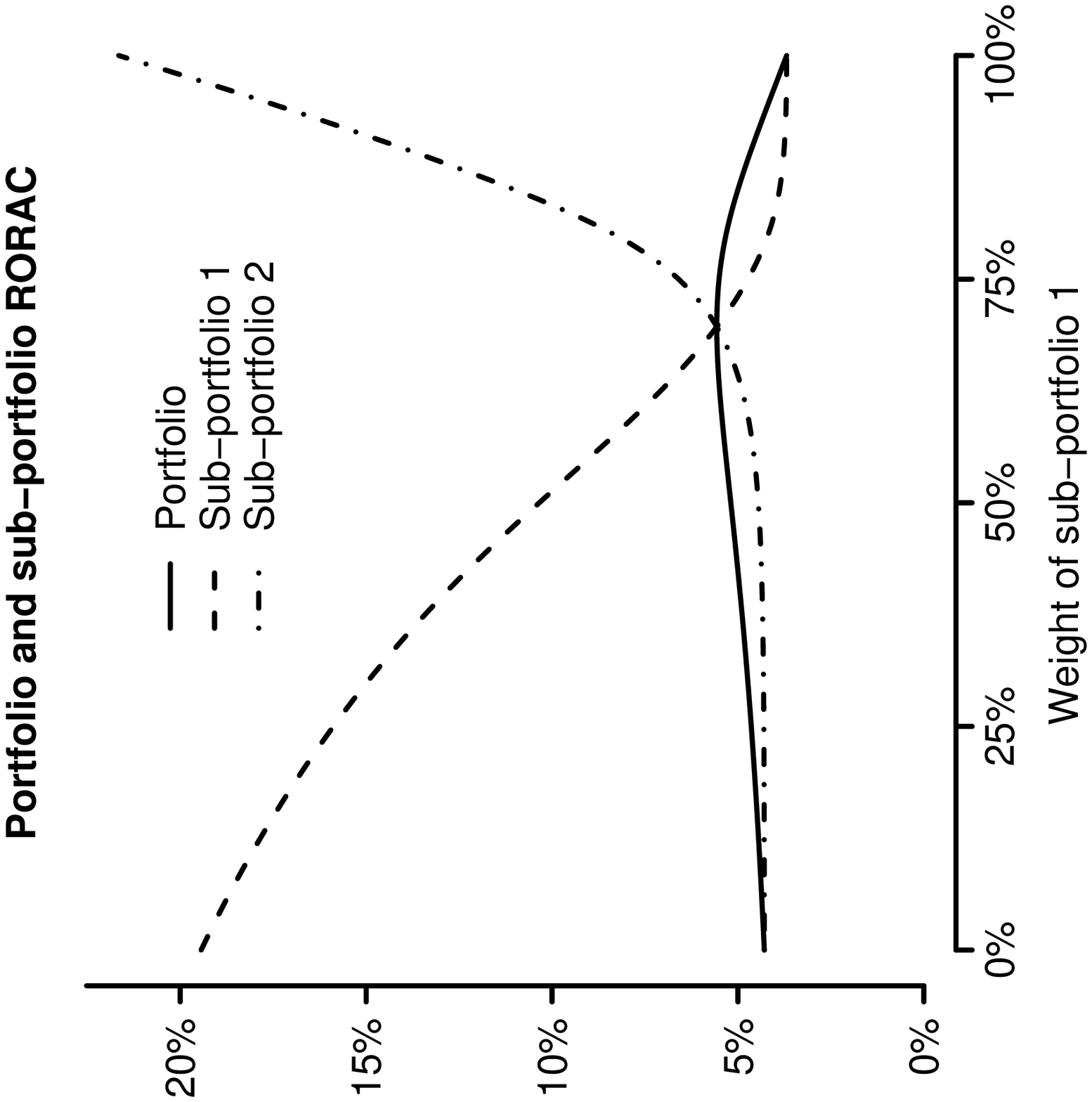}}
\end{turn}
\fi
\end{center}
\end{figure}
Figure \ref{fig:RORAC} demonstrates the concept of Definition \ref{de:desirable} and Proposition \ref{pr:determ} with a two-sub-portfolios example\footnote{%
The risk measure chosen for this example is unexpected loss based on Value-at-Risk (see \eqref{eq:ULVaR}).}. As long as $\mathrm{RORAC}(X_1\,|\,X) \not= \mathrm{RORAC}(X)$ or $\mathrm{RORAC}(X_2\,|\,X) \not= \mathrm{RORAC}(X)$, the portfolio-wide RORAC (when considered as function of the weight of the first sub-portfolio) does not reach its optimum. Only at a weight of about 70\% of sub-portfolio~1, when all the three RORAC-curves intersect, the portfolio-wide RORAC obtains its maximum because no further improvement by shifting the portfolio-weights is possible.

What about the full allocation property of Definition \ref{de:desirable}? Assume that the function $f_\rho$ corresponding to the risk measure $\rho$ is continuously differentiable. Then, by Euler's theorem on homogeneous functions (see Theorem \ref{th:euler} in Appendix \ref{sec:eulertheorem}), $f_\rho$ satisfies the equation
\begin{equation}\label{eq:homspec}
    f_\rho(u)\ = \sum_{i=1}^n u_i\,\frac{\partial\,f_\rho(u)}{\partial\,u_i}
\end{equation}
for all $u$ in its range of definition if and only if it is homogeneous of degree 1 (cf.\ Definition
\ref{de:homo}). Consequently, for the risk contributions to risk measures $\rho$ with continuously differentiable $f_\rho$ the two properties of Definition \ref{de:desirable} can obtain at the same time if and only if the risk measure is homogeneous of degree 1. The risk contributions are then uniquely determined by \eqref{eq:deriv}.

\begin{remark}\label{de:euleralloc}
If $\rho$ is a risk measure which is homogeneous of degree 1 (in the sense of Definition \ref{de:homo}), then risk contributions according to \eqref{eq:deriv} are called \emph{Euler contributions}. Euler contributions satisfy both properties of Definition \ref{de:desirable}, i.e.\ they are RORAC compatible and add up to portfolio-wide risk. The process of assigning capital to assets or sub-portfolios by calculating Euler contributions is called \emph{Euler allocation.}
\end{remark}

The use of the
Euler allocation principle was justified by several authors with
different reasonings:
\begin{itemize}
    \item \citet{Patriketal} argued from a practitioner's view emphasizing
    mainly the fact that the risk contributions according to the Euler principle
    by \eqref{eq:deriv} naturally add up to the portfolio-wide economic capital.
    \item \citet{Litterman96} and \citet[][as mentioned above]{Tasche1999} pointed out that the Euler
    principle is fully compatible with economically sensible portfolio diagnostics
    and optimization.
    \item \citet{Denault01} derived the Euler principle
    by game-theoretic considerations.
    \item In the context of capital allocation for insurance companies, \cite{MyersRead}
    argued that applying the Euler principle to the expected ``default value'' (essentially
    $\mathrm{E}[\max(X, 0)]$) of the insurance portfolio is most appropriate for deriving
    line-by-line surplus requirements.
    \item \citet{Kalkbrener05} presented an axiomatic approach to
    capital allocation and risk contributions. One of his axioms requires
    that risk contributions do not exceed the corresponding stand-alone risks.
    From this axiom in connection with more technical conditions, in the context
    of sub-additive and positively homogeneous risk measures, \citeauthor{Kalkbrener05}
    concluded that the Euler principle is the only allocation principle to be
    compatible with the ``diversification''-axiom (see also \citeauthor{Kalkbreneretal04},
    \citeyear{Kalkbreneretal04}; \citeauthor{Tasche2002}, \citeyear{Tasche2002}; and
    Section \ref{se:subadd_contributions} below).
    \item More recently, the Euler allocation was criticized for not being compatible with
    the decentralized risk management functions of large financial institutions \citep{Schwaiger2006}.
    \citet{Grundl2007} even find that capital allocation is not needed at all for insurance companies.
\end{itemize}

\subsection{Contributions to sub-additive risk measures}
\label{se:subadd_contributions}

As risk hedging by diversification plays a major role for portfolio management, we briefly recall the observations by \citet{Kalkbrener05} and \citet{Tasche1999} on the relation between the Euler allocation principle and diversification.

It is quite common to associate risk measures that reward portfolio diversification with the so-called sub-additivity property \citep{ADEH99}. A risk measure $\rho$ is \emph{sub-additive} if it satisfies
\begin{equation}\label{eq:subadd}
    \rho(X+Y) \ \le \ \rho(X) + \rho(Y)
\end{equation}
for any random variables $X$, $Y$ in its range of definition. Assume the setting of Section \ref{se:euler} and that $\rho$ is a risk  measure that is both homogeneous of degree 1 and sub-additive. By Corollary \ref{co:sub}, then the function $f_\rho$ that corresponds to $\rho$ via \eqref{eq:fct_risk} fulfills the inequality
\begin{subequations}
\begin{equation}\label{eq:divers}
    \sum_{i=1}^n u_i\,\frac{\partial\,f_\rho(u+v)}{\partial\,u_i}\ \le \ f_\rho(u), \quad i=1,\ldots,n.
\end{equation}
With $u = (0, \ldots, 0, 1, 0, \ldots, 0)$ (1 at $i$-th position) and $v = (1, \ldots, 1) - u$, \eqref{eq:divers} implies
\begin{equation}\label{eq:contribdivers}
    \rho_{\mathrm{Euler}}(X_i\,|\,X)\ \le\ \rho(X_i), \quad i = 1,\ldots,n,
\end{equation}
where $\rho_{\mathrm{Euler}}(X_i\,|\,X)$ is defined by \eqref{eq:deriv}.
\end{subequations}
Hence, if risk contributions to a homogeneous and sub-additive risk measure are calculated as Euler contributions, then
the contributions of single assets will never exceed the assets' stand-alone risks. In particular, risk contributions of credit assets
cannot become larger than the face values of the assets.

Actually, Corollary \ref{co:sub} shows that, for continuously differentiable and risk measures $\rho$ homogeneous of degree 1, property \eqref{eq:contribdivers} for the Euler contributions and sub-additivity of the risk measure are equivalent. This is of particular relevance for credit risk portfolios where violations of the sub-additivity property are rather observed as violations of \eqref{eq:contribdivers} than as violations of \eqref{eq:subadd} \citep[see][]{Kalkbreneretal04}.

Recall the notion of the so-called \emph{marginal risk contribution}\footnote{%
This methodology is also called \emph{with-without principle} by some authors.
} for determining the capital
required by an individual business, asset, or sub-portfolio. Formally, the marginal risk
contribution
$\rho_{\mathrm{marg}}(X_i\,|\,X)$ of asset $i,\ i =1, \ldots,n$, is defined by
\begin{equation}
\rho_{\mathrm{marg}}(X_i\,|\,X) \ =\  \rho(X) - \rho(X-X_i),
  \label{g503}
\end{equation}
i.e.~by the difference of the portfolio risk with asset $i$ and the
portfolio risk without asset $i$. In the case of continuously differentiable
and sub-additive risk measures that are homogeneous of degree 1, it can be shown that
marginal risk contributions are always smaller than the corresponding Euler contributions
\citep[][Proposition 2]{Tasche2004a}.

\begin{proposition}\label{pr:p_neu}
Let $\rho$ be  a sub-additive
and continuously differentiable risk
measure that is homogeneous of degree 1. Then the  marginal risk contributions $\rho_{\mathrm{marg}}(X_i\,|\,X)$
as defined by \eqref{g503} are smaller than the corresponding Euler contributions, i.e.
\begin{subequations}
\begin{equation}
  \label{eq:under}
  \rho_{\mathrm{marg}}(X_i\,|\,X) \ \le\ \rho_{\mathrm{Euler}}(X_i\,|\,X).
\end{equation}
In particular, the sum of the marginal risk contributions underestimates total risk, i.e.
\begin{equation}\label{eq:underest}
    \sum_{i=1}^n \rho_{\mathrm{marg}}(X_i\,|\,X)\ =\
    \sum_{i=1}^n \bigl(\rho(X) - \rho(X-X_i)\bigr)\ \le\ \rho(X).
\end{equation}
\end{subequations}
\end{proposition}

As a work-around for the problem that marginal risk contributions do not satisfy the
full allocation property, sometimes marginal risk contributions are defined as
\begin{equation}
\rho^\ast_{\mathrm{marg}}(X_i\,|\,X) \ =\  \frac{\rho_{\mathrm{marg}}(X_i\,|\,X)}
{\sum_{j=1}^n \rho_{\mathrm{marg}}(X_j\,|\,X)}\,\rho(X).
  \label{g503ast}
\end{equation}
This way, equality in \eqref{eq:underest} is forced. In general, however, marginal
risk contributions according to \eqref{g503ast} do not fulfil the RORAC compatibility
property from Definition \ref{de:desirable}.


\subsection{Measuring concentration and diversification}
\label{se:concentration}

In \citet[][paragraph 770]{BaselAccord} the Basel Committee on Banking Supervision
states: ``A risk concentration is any single exposure or group of exposures with the potential
to produce losses large enough (relative to a bank's capital, total assets, or overall risk level)
to threaten a bank's health or ability to maintain its core operations. Risk concentrations are
arguably the single most important cause of major problems in banks.'' In \citet[][paragraph 774]{BaselAccord}
the Basel Committee then explains: ``A bank's framework for managing credit risk concentrations should be clearly
documented and should include a definition of the credit risk concentrations relevant to the
bank and how these concentrations and their corresponding limits are calculated.'' We demonstrate in this section that the Euler allocation as introduced in Section \ref{se:contributions} is particularly well suited for calculating concentrations.

The concept of concentration index \citep[following][]{Tasche2006} we will introduce is based on the idea that the actual risk of a portfolio should be compared to an appropriate worst-case risk of the portfolio in order to be able to identify risk concentration.
It turns out that ``worst-case risk'' can be adequately expressed as maximum dependence of the random variables
the portfolio model is based on.
In actuarial science, the concept of co-monotonicity
    is well-known as it supports easy and reasonably conservative representations of dependence structures
    \citep[see, e.g.,][]{Dhaeneetal}.
    Random variables $V$ and $W$ are called
    co-monotonic if they can be represented as non-decreasing functions of a third
    random variable $Z$, i.e.
    \begin{subequations}
    \begin{equation}\label{eq:co-mono}
    V = h_V(Z) \quad \text{and}\quad W = h_W(Z)
\end{equation}
for some non-decreasing functions $h_V, h_W$. As co-monotonicity is
implied if $V$ and $W$ are correlated with correlation coefficient
1, it generalizes the concept of linear dependence. A risk measure
$\rho$ is called \emph{co-monotonic additive} if for any co-monotonic
random variables $V$ and $W$
\begin{equation}\label{eq:add}
    \rho(V+W)\ = \ \rho(V) + \rho(W).
\end{equation}
\end{subequations}
Thus co-monotonic additivity can be interpreted as a specification of worst case scenarios
when risk is measured by a sub-additive (see \eqref{eq:subadd}) risk measure:
nothing worse can occur than co-monotonic random variables --
which seems quite natural\footnote{%
For standard deviation based risk measures \eqref{eq:add} obtains if and only if
$V$ and $W$ are fully linearly correlated (i.e.\ correlated with correlation coefficient 1).
Full linear correlation implies co-monotonicity but co-monotonicity does not imply full
correlation. Standard deviation based worst-case scenarios, therefore, might be considered
``non-representative''.
}.
These observations suggest the first part of the following definition.

\begin{definition}\label{de:factor}
Let $X_1, \ldots, X_n$ be real-valued random variables and let $X
= \sum_{i=1}^n X_i$. If $\rho$ is a risk measure such that
$\rho(X), \rho(X_1), \ldots, \rho(X_n)$ are defined, then
\begin{subequations}
\begin{equation}\label{eq:port}
    \mathrm{DI}_{\rho}(X) \ = \
    \frac{\rho(X)}
    {\sum_{i=1}^n \rho(X_i)}
\end{equation}
denotes the \emph{diversification index} of portfolio $X$ with
respect to
the risk measure $\rho$.\\
If Euler risk contributions of $X_i$ to $\rho(X)$ in the sense of Remark
\ref{de:euleralloc} exist, then the ratio
\begin{equation}\label{eq:single}
    \mathrm{DI}_{\rho}(X_i\,|\,X) \ = \
    \frac{\rho_{\mathrm{Euler}}(X_i\,|\,X)}{\rho(X_i)}
\end{equation}
\end{subequations}
with $\rho_{\mathrm{Euler}}(X_i\,|\,X)$ being defined by
\eqref{eq:deriv} denotes the \emph{marginal
diversification index} of sub-portfolio
$X_i$ with respect to the risk measure $\rho$.
\end{definition}
Note that without calling the concept ``diversification index'',
\citet{MemmelWehn} calculate a diversification index for the
German supervisor's market price risk portfolio. \citet{Garcia04} use the
diversification indices as defined here
for a representation of portfolio risk as a ``diversification
index''-weighted sum of stand-alone risks.

\begin{remark}\label{rm:index}
Definition \ref{de:factor} is most useful when the risk measure $\rho$ under consideration
is homogeneous of degree 1, sub-additive, and co-monotonic additive. Additionally, the function $f_\rho$
associated with $\rho$ via \eqref{eq:homspec} should be continuously differentiable. Expected shortfall,
as considered in Section \ref{se:ES}, enjoys homogeneity, sub-additivity, and co-monotonic additivity. Its associated
function is continuously differentiable under moderate assumptions on the joint distribution
of the variables $X_i$ \citep[cf.][]{Tasche1999, Tasche2002}.

Assume that $\rho$ is a risk measure that has these four properties. Then
\begin{itemize}
  \item[(i)] by sub-additivity, $\mathrm{DI}_{\rho}(X) \, \le\, 1$
  \item[(ii)] by co-monotonic additivity, $\mathrm{DI}_{\rho}(X) \, \approx\, 1$ indicates that $X_1, \ldots, X_n$
  are ``almost'' co-monotonic (i.e.\ strongly dependent)
  \item[(iii)] by \eqref{eq:contribdivers}, $\mathrm{DI}_{\rho}(X_i\,|\,X) \, \le\, 1$
  \item[(iv)] by Proposition \ref{pr:determ}, $\mathrm{DI}_{\rho}(X_i\,|\,X) < \mathrm{DI}_{\rho}(X)$ implies that
  there is some $\epsilon_i > 0$ such that $\mathrm{DI}_{\rho}(X+h\,X_i) < \mathrm{DI}_{\rho}(X)$ for $0 < h < \epsilon_i$.
\end{itemize}
\end{remark}
With regard to conclusion (ii) in Remark \ref{rm:index}, a portfolio with a diversification index close to 100\% might be
considered to have high risk concentration whereas a portfolio with a low diversification index might be considered
well diversified. However, it is not easy to specify how far from 100\% a diversification index should be for
implying that the portfolio is well diversified.

Conclusion (iv) in Remark \ref{rm:index} could be more useful for
such a distinction between concentrated and diversified portfolios, because marginal diversification indices indicate
rather diversification potential than ``absolute'' diversification. In this sense, a portfolio with high unrealized diversification
potential could be regarded as concentrated.

\section{Practice}
\label{sec:practice}

On principle, formula \eqref{eq:deriv} can be applied directly for calculating Euler risk contributions. It has turned out, however, that for some popular families of risk measures it is possible to derive closed-form expressions for the involved derivative. In this section, results are presented for standard deviation based risk measures (Section \ref{se:stddev}), Value-at-Risk (Section \ref{se:VaR}), and Expected Shortfall (Section \ref{se:ES})\footnote{%
With respect to other classes of risk measures see, e.g., \citet{Fischer} for a discussion of derivatives of one-sided moment measures and \citet{Tasche2002} for a suggestion of how to apply \eqref{eq:VaR_contrib} to spectral risk measures.}.
In Section \ref{se:sample}, we discuss in some detail how to implement \eqref{eq:deriv} in the case of VaR when the underlying distribution has to be inferred from a sample. A related but somehow different application of Euler's theorem to the calculation of single name contributions to the expected loss of CDO tranches is discussed in Section \ref{se:CDO}. Section \ref{se:nl_impact} presents another application of Euler allocation to the measurement of risk impacts on non-linear portfolios. 
In the following, the notation of Section \ref{se:euler} is adopted.

\subsection{Standard deviation based risk measures}
\label{se:stddev}

We consider here the family of risk measures $\sigma_c$, $c > 0$ given by
\begin{equation}\label{eq:stddev}
    \sigma_c(X)\ =\ c\,\sqrt{\mathrm{var}[X]}\ = \ c\,\sqrt{\mathrm{E}[(X-\mathrm{E}[X])^2]}.
\end{equation}
It is common to choose the constant $c$ in such a way that
\begin{equation}\label{eq:restrict}
    \mathrm{P}[X \le \mathrm{E}[X] - \sigma_c(X)]\ \le\ 1-\alpha,
\end{equation}
where $\alpha$ denotes some -- usually large -- probability (like 99\% or 99.95\%). Sometimes this is done assuming that
$X$ is normally distributed. A robust alternative would be an application of the one-tailed Chebychev-inequality:
\begin{equation}\label{eq:chebychev}
    \mathrm{P}[X \le \mathrm{E}[X] - \sigma_c(X)]\ \le \ \frac{1}{1+c^2}.
\end{equation}
Solving $1/(1+c^2) = 1-\alpha$ for $c$ will then ensure that \eqref{eq:restrict} obtains for all $X$ with finite variance.
Note that, however, this method for determining $c$ will yield much higher values of $c$ than the method based on a
normal-distribution assumption.

The risk measures $\sigma_c$, $c > 0$ are homogeneous of degree 1 and sub-additive, but not co-monotonic additive.
For $X = \sum_{i=1}^n X_i$ the Euler contributions according to Remark \ref{de:euleralloc} can readily be calculated by
differentiation:
\begin{subequations}
\begin{equation}\label{eq:contrib_stddev}
    \sigma_c(X_i\,|\,X)\ =\ c\,\frac{\mathrm{cov}[X_i,\,X]}{\sqrt{\mathrm{var}[X]}}.
\end{equation}
In case that $X_i$ is given as $g_i - L_i$ (cf.\ \eqref{eq:XL}), we have $\sigma_c(X) = \sigma_c(L)$ and
\begin{equation}\label{eq:contrib_stddev_L}
    \sigma_c(X_i\,|\,X)\ =\ c\,\frac{\mathrm{cov}[L_i,\,L]}{\sqrt{\mathrm{var}[L]}}.
\end{equation}
\end{subequations}

\subsection{Value-at-Risk}
\label{se:VaR}

For any real-valued random variable $Y$ and $\gamma \in (0,1)$ define the $\gamma$-quantile of $Y$ by
\begin{subequations}
\begin{equation}\label{eq:quantile}
    q_\gamma(Y)\ =\ \min\{y: \mathrm{P}[Y \le y] \ge \gamma\}.
\end{equation}
If $Y$ has a strictly increasing and continuous distribution function $F(y) = \mathrm{P}[Y\le y]$, quantiles
of $Y$ can be expressed by the inverse function of $F$:
\begin{equation}\label{eq:Fquantile}
    q_\gamma(Y)\ =\ F^{-1}(\gamma).
\end{equation}
\end{subequations}
For a portfolio-wide profit/loss variable $X = \sum_{i=1}^n X_i$ the \emph{Value-at-Risk (VaR)} of $X$ at confidence level $\alpha$ ($\alpha$ usually
close to 1) is defined as the $\alpha$-quantile of $-X$:
\begin{equation}\label{eq:VaR}
    \mathrm{VaR}_\alpha(X)\ =\ q_\alpha(-X).
\end{equation}
VaR as a risk measure is homogeneous of degree 1 and co-monotonic additive but not in general sub-additive.
Under some smoothness conditions (see \citeauthor{GL00}, \citeyear{GL00}, or
\citeauthor{Tasche1999}, \citeyear{Tasche1999}, Section 5.2), a general formula can be derived for
the Euler contributions to $\mathrm{VaR}_\alpha(X)$ according to Remark \ref{de:euleralloc}. These smoothness conditions,
in particular, imply that $X$ has a density. The formula for the Euler VaR-contributions reads
\begin{subequations}
\begin{equation}\label{eq:VaR_contrib}
    \mathrm{VaR}_\alpha(X_i\,|\,X)\ =\ - \mathrm{E}[X_i\,|\,X = - \mathrm{VaR}_\alpha(X)],
\end{equation}
where $\mathrm{E}[X_i\,|\,X]$ denotes the \emph{conditional expectation} of $X_i$ given $X$.
In case that $X_i$ is given as $g_i - L_i$ (cf.\ \eqref{eq:XL}), we have $\mathrm{VaR}_\alpha(X) = q_\alpha(L) - \sum_{i=1}^n g_i$ and
\begin{equation}\label{eq:VaR_contrib_L}
    \mathrm{VaR}_\alpha(X_i\,|\,X)\ =\ \mathrm{E}[L_i\,|\,L = q_\alpha(L)] - g_i.
\end{equation}
\end{subequations}
Often, it is not VaR itself that is of interest but rather \emph{Unexpected Loss:}
\begin{subequations}
\begin{equation}\label{eq:ULVaR}
    \mathrm{UL}_{\mathrm{VaR}, \alpha}(X)\ =\ \mathrm{VaR}_\alpha(X - \mathrm{E}[X])\ = \ \mathrm{VaR}_\alpha(X) + \mathrm{E}[X].
\end{equation}
In terms of $X_i = g_i - L_i$, Equation \eqref{eq:ULVaR} reads
\begin{equation}\label{eq:ULVaR_L}
    \mathrm{UL}_{\mathrm{VaR}, \alpha}(X)\ =\ \mathrm{VaR}_\alpha(\mathrm{E}[L]-L)\ = \ q_\alpha(L) - \mathrm{E}[L].
\end{equation}
\end{subequations}
Having in mind that the Euler contribution of $X_i$ to $\mathrm{E}[X]$ is obviously $\mathrm{E}[X_i]$, the formulae for
the Euler contributions to $\mathrm{UL}_{\mathrm{VaR}, \alpha}(X)$ are obvious from Equations \eqref{eq:VaR_contrib}, \eqref{eq:VaR_contrib_L},
\eqref{eq:ULVaR}, and \eqref{eq:ULVaR_L}.

In general, the conditional expectation of $X_i$ given $X$ cannot easily be calculated or
estimated. For some exceptions from this observation see \citet{Tasche2004} or \citet{Tasche2006}.
As the conditional expectation of $X_i$ given $X$ can be interpreted as the \emph{best prediction} of $X_i$ by $X$ in a least squares
context, approximation of $\mathrm{VaR}_\alpha(X_i\,|\,X)$ by \emph{best linear predictions} of $X_i$ by $X$ has been proposed.
Linear approximation of the right-hand side of \eqref{eq:VaR_contrib} by $X$ and a constant yields
\begin{equation}\label{eq:approx_contrib}
\mathrm{VaR}_\alpha(X_i\,|\,X)\ \approx \ \frac{\mathrm{cov}%
[X_i,\,X]}{\mathrm{var}[X]}\,\mathrm{UL}_{\mathrm{VaR}, \alpha}(X) - \mathrm{E}[X_i].
\end{equation}
The approximation in \eqref{eq:approx_contrib} can be improved by additionally admitting quadratic or other non-linear transformations
of $X$ as regressors \citep[cf.][Section 5]{TascheTibiletti}. All such regression-based approximate Euler contributions to VaR
satisfy the full allocation property of Definition \ref{de:desirable} but are not RORAC compatible. See Section \ref{se:sample} for an approach to the estimation of \eqref{eq:VaR_contrib} that yields RORAC compatibility.

\subsection{Expected shortfall}
\label{se:ES}

For a portfolio-wide profit/loss variable $X = \sum_{i=1}^n X_i$ the \emph{Expected Shortfall (ES)}\footnote{%
The denotation ``Expected Shortfall'' was proposed by \citet{Acerbi&Tasche}. A common
alternative denotation is ``Conditional Value-at-Risk (CVaR)'' that was suggested by
\citet{Rockafellar&Uryasev}.
} of $X$ at confidence level $\alpha$ ($\alpha$ usually
close to 1) is defined as an average of VaRs of $X$ at level $\alpha$ and higher:
\begin{subequations}
\begin{equation}\label{eq:ES}
    \mathrm{ES}_\alpha(X)\ =\ \frac 1 {1-\alpha} \int_\alpha^1 \mathrm{VaR}_u(X)\,d u.
\end{equation}
ES as a risk measure is homogeneous of degree 1, co-monotonic additive and sub-additive.
Under some smoothness conditions \citep[see][Section.5.3]{Tasche1999}, a general formula can be derived for
the Euler contributions to $\mathrm{ES}_\alpha(X)$ according to Remark \ref{de:euleralloc}. These smoothness conditions,
in particular, imply that $X$ has a density. In that case, ES can equivalently be written as
\begin{equation}\label{eq:ESequi}
    \mathrm{ES}_\alpha(X)\ =\ - \mathrm{E}[X\,|\,X \le - \mathrm{VaR}_\alpha(X)].
\end{equation}
\end{subequations}
The formula for the Euler ES-contributions reads
\begin{subequations}
\begin{align}
    \mathrm{ES}_\alpha(X_i\,|\,X)&\ =\ - \mathrm{E}[X_i\,|\,X \le - \mathrm{VaR}_\alpha(X)]\notag\\
    &\ = \ - (1-\alpha)^{-1} \mathrm{E}[X_i\,\mathbf{1}_{\{X \le - \mathrm{VaR}_\alpha(X)\}}].\label{eq:ES_contrib}
\end{align}
Note that $\mathrm{E}[X_i\,|\,X \le - \mathrm{VaR}_\alpha(X)]$, in contrast to $\mathrm{E}[X_i\,|\,X]$ from
\eqref{eq:VaR_contrib}, is an \emph{elementary conditional expectation} because the conditioning event
has got a  positive probability to occur.
In case that $X_i$ is given as $g_i - L_i$ (cf.\ \eqref{eq:XL}), we have $\mathrm{ES}_\alpha(X) =
        \frac 1 {1-\alpha} \int_\alpha^1 q_u(L)\,du - \sum_{i=1}^n g_i$ and
\begin{equation}\label{eq:ES_contrib_L}
    \mathrm{ES}_\alpha(X_i\,|\,X)\ =\ \mathrm{E}[L_i\,|\,L \ge q_\alpha(L)] - g_i.
\end{equation}
\end{subequations}
Often, it is not ES itself that is of interest but rather \emph{Unexpected Loss:}
\begin{subequations}
\begin{equation}\label{eq:ULES}
    \mathrm{UL}_{\mathrm{ES}, \alpha}(X)\ =\ \mathrm{ES}_\alpha(X - \mathrm{E}[X])\ = \ \mathrm{ES}_\alpha(X) + \mathrm{E}[X].
\end{equation}
In terms of $X_i = g_i - L_i$, Equation \eqref{eq:ULES} reads
\begin{equation}\label{eq:ULES_L}
    \mathrm{UL}_{\mathrm{ES}, \alpha}(X)\ =\ \mathrm{ES}_\alpha(\mathrm{E}[L]-L)\ = \ \frac 1 {1-\alpha} \int_\alpha^1 q_u(L)\,du - \mathrm{E}[L].
\end{equation}
\end{subequations}
Having in mind that the Euler contribution of $X_i$ to $\mathrm{E}[X]$ is obviously $\mathrm{E}[X_i]$, the formulae for
the Euler contributions to $\mathrm{UL}_{\mathrm{ES}, \alpha}(X)$ are obvious from Equations \eqref{eq:ES_contrib}, \eqref{eq:ES_contrib_L},
\eqref{eq:ULES}, and \eqref{eq:ULES_L}.

In contrast to Euler contributions to VaR, thanks to  representation \eqref{eq:ES_contrib}, estimation of Euler contributions to
ES is quite straightforward.

\subsection{Risk measures for sample data}
\label{se:sample}

In most circumstances, portfolio loss distributions cannot be calculated analytically but have to be estimated from simulated or historical sample data. For a portfolio of $n$ assets, as specified in Section \ref{se:euler}, the sample data might be given as $n$-dimensional points
\begin{equation}\label{eq:sample}
 (x_{1,1}, \ldots, x_{n,1}), \ldots, (x_{1,N}, \ldots, x_{n,N}),
\end{equation}
where $x_{i,k}$ denotes the profit/loss of asset $i$ in the $k$-th observation (of $N$). Each data point
$(x_{1,k}, \ldots, x_{n,k})$ would be interpreted as a realization of the profit/loss random vector $(X_1, \ldots, X_n)$. The portfolio-wide profit/loss in the $k$-th observation would then be obtained as $x_k = \sum_{i=1}^n x_{i,k}$. If the sample is large and the observations can be assumed independent, by the law of large numbers the empirical measure
\begin{equation}\label{eq:emp_measure}
\begin{split}
    \widehat{\mathrm{P}}_N(A) &\ =\ \frac 1 N \sum_{k=1}^N \delta_A(x_{1,k}, \ldots, x_{n,k}), \quad A \subset \mathbb{R}^n \ \text{measurable},\\
    \delta_A(x_{1,k}, \ldots, x_{n,k}) &\ =\ \left\{
      \begin{array}{ll}
        1, & \hbox{if $(x_{1,k}, \ldots, x_{n,k})\in A$;} \\
        0, & \hbox{otherwise}
      \end{array}
    \right.
\end{split}
\end{equation}
will approximate the joint distribution $\mathrm{P}[(X_1, \ldots, X_n) \in A]$ of the assets' profits/losses $X_i$.

Denote by $(\widehat{X}_1, \ldots, \widehat{X}_n)$ the profit/loss random vector under the empirical measure $\widehat{\mathrm{P}}_N$. Let $\widehat{X} = \sum_{i=1}^n \widehat{X}_i$ be the portfolio-wide profit/loss under
the empirical measure $\widehat{\mathrm{P}}_N$. In the cases of standard deviation based risk measures and Expected Shortfall, then \emph{statistically consistent} estimators for the risk contributions according to \eqref{eq:contrib_stddev} and \eqref{eq:ES_contrib} respectively can be obtained by simply substituting $\widehat{X}, \widehat{X}_i$ for $X, X_i$. This naive approach does not work for Euler contributions to VaR according to \eqref{eq:VaR_contrib}, if $X$ has a continuous distribution. In this case, smoothing of the empirical measure
(\emph{kernel estimation}) as described in the following can help.

Let $\xi$ be a random variable which is independent of $(\widehat{X}_1, \ldots, \widehat{X}_n)$ and has a continuous density (or \emph{kernel}) $\varphi$ ($\xi$ standard normal would be a good and convenient choice). Fix some $b>0$ (the \emph{bandwidth}). Then $\widehat{X} + b\,\xi$ has the density
\begin{equation}\label{eq:dens}
        \widehat{f}_b(x)\ = \widehat{f}_{b, x_1, \ldots, x_N}(x)\ =\ \frac1{b\,N} \sum_{k=1}^N \varphi\bigl(\tfrac{x-x_k}b\bigr).
\end{equation}
Actually, \eqref{eq:dens} represents the well-known \emph{Rosenblatt-Parzen} estimator of the density of $X$.
If the kernel $\varphi$ and the density of $X$  are appropriately ``smooth'' \citep[see, e.g.,][Theorem 2.5 for details]{Pagan99},
it can be shown for
$b_N \to 0$, $b_N\,N\to\infty$ that $\widehat{f}_{b_N}(x)$ is a point-wise mean-squared consistent
estimator of the density of $X$. Whereas the choice of the kernel $\varphi$, subject to some mild conditions, is not too important for the efficiency of the Rosenblatt-Parzen estimator, the appropriate choice of the bandwidth is crucial.
\emph{Silverman}'s rule of thumb \citep[cf.\ (2.52) in][]{Pagan99}
\begin{equation}\label{eq:Silverman}
    b\ =\ 0.9\,\min(\sigma, R/1.34)\,N^{-1/5}
\end{equation}
is known to work quite well in many circumstances.
In \eqref{eq:Silverman}, $\sigma$ and $R$ denote the standard deviation and the interquartile range
respectively of the sample $x_1, \ldots, x_N$. In the case where $X$ is not heavy-tailed (as is the case for credit portfolio loss distributions), another quite promising and easy-to-implement method for the bandwidth selection is \emph{Pseudo Likelihood Cross Validation}. See, for instance, \citet{Turlach} for an overview of this and other selection methods.

In the context of the Euler allocation, a major advantage with the kernel estimation approach is that the smoothness conditions from \citet[][Section 5.2]{Tasche1999} are obtained for $\xi, \widehat{X}_1, \ldots, \widehat{X}_n$. As a consequence, it follows from \citet[][Lemma 5.3]{Tasche1999} that
\begin{align}
    \mathrm{VaR}_\alpha(X_i\,|\,X) &\ \approx\ \mathrm{VaR}_\alpha(\widehat{X}_i\,|\,\widehat{X}+b\,\xi)\notag\\[1ex]
    &\ =\ \frac{d\, \mathrm{VaR}_\alpha}{d\,h}(\widehat{X} + b\,\xi +h\,\widehat{X}_i)\bigl|_{h=0}\notag\\[1ex]
    &\ =\ -\,\mathrm{E}[\widehat{X}_i\,|\,\widehat{X} +b\,\xi=-\mathrm{VaR}_\alpha(\widehat{X} +b\,\xi)]\notag\\[1ex]
    &\ =\ -\,\frac{\sum\nolimits_{k=1}^N x_{i,k}\,\varphi\bigl(\tfrac{-\mathrm{VaR}_\alpha(\widehat{X}
        +b\,\xi)-x_k}b\bigr)}
      {\sum\nolimits_{k=1}^N \varphi\bigl(\tfrac{-\mathrm{VaR}_\alpha(\widehat{X}
        +b\,\xi)-x_k}b\bigr)}.\label{eq:smooth}
\end{align}
The right-hand side of \eqref{eq:smooth} is just the \emph{Nadaraya-Watson} kernel estimator of $\mathrm{E}[X_i\,|\,X = -\mathrm{VaR}_\alpha(X)]$. Note that it is clear from the derivation of \eqref{eq:smooth} in the context of the empirical measure \eqref{eq:emp_measure} that the bandwidth  $b$ in \eqref{eq:smooth} should be the same as in \eqref{eq:dens}. By construction, we obtain for the sum of the approximate Euler contributions to VaR according to \eqref{eq:smooth}
\begin{align}
    \sum_{i=1}^n \mathrm{VaR}_\alpha(\widehat{X}_i\,|\,\widehat{X}+b\,\xi) &\ =\ \mathrm{VaR}_\alpha(\widehat{X}+b\,\xi) - \mathrm{VaR}_\alpha(\xi\,|\,\widehat{X}+b\,\xi)\notag\\
&\ =\ \mathrm{VaR}_\alpha(\widehat{X}+b\,\xi) + \mathrm{E}[\xi\,|\,
        \widehat{X} +b\,\xi=-\mathrm{VaR}_\alpha(\widehat{X} +b\,\xi)].\label{eq:VaR_sum}
\end{align}
The sum of the approximate Euler contributions therefore differs from natural estimates of $\mathrm{VaR}\alpha(X)$ such as $\mathrm{VaR}_\alpha(\widehat{X})$ or $\mathrm{VaR}_\alpha(\widehat{X}+b\,\xi)$. Practical experience shows that the difference tends to be small. Some authors \citep[e.g.][]{EpperleinSmillie} suggest to account for this difference by an appropriate multiplier. Another way to deal with the issue could be to take the left-hand side of \eqref{eq:VaR_sum} as an estimate for $\mathrm{VaR}_\alpha(X)$.

\citet{Yamai02} found that estimates for ES and VaR Euler contributions are very volatile. See \citet{Glasserman2005} and \citet{Merino04} for methods to tackle this problem for ES by importance sampling and \citet{Tasche2007} for an approach to Euler VaR contribution estimation by importance sampling.

\subsection{Expected loss decomposition for CDO tranches}
\label{se:CDO}

In the case of a credit portfolio with loss variable $L \ge 0$ given as
\begin{subequations}
\begin{align}\label{eq:L}
    L & = \sum_{i=1}^n L_i, \\
    \intertext{or in a dynamic setting with exposure sizes $u_i$}
    L(u) & = \sum_{i=1}^n u_i\,L_i, \label{eq:Lu}
\end{align}
\end{subequations}
the loss variables $L_i \ge 0$ for single assets represent a natural vertical decomposition of the
portfolio-wide loss $L$. If the portfolio serves as the underlying of a collateralized debt obligation (CDO), the normalized (i.e.\ percentage) portfolio-wide loss $0 \le L \le 1$ is decomposed into tranches
in a horizontal manner, via
\begin{subequations}
\begin{align}\label{eq:L_CDO}
    L & = \sum_{j=1}^m Y_j \\
    \intertext{with}
    Y_j & = \min(L, c_j) - \min(L, c_{j-1}), \quad j = 1, \ldots, m\label{eq:Y}
\end{align}
where $0 = c_0 < c_1 < \ldots < c_{m-1} < c_m = 1$, $m \ge 2$ denote the so-called \emph{credit enhancement levels} of the CDO\footnote{%
This is actually a simplifying approach, ignoring in particular timing effects \citep[cf.][Section 4.4]{Bluhm2003}.}.
\end{subequations}
Hence, the random variable $Y_j$ is associated with the loss suffered with the $j$-th tranche of the CDO and is characterized by the credit enhancement levels $c_{j-1}$ and $c_j$.

According to \eqref{eq:L} and \eqref{eq:L_CDO}, in particular, the portfolio expected loss $\mathrm{E}[L]$ can be decomposed in two ways, namely as
\begin{subequations}
\begin{align}
\mathrm{E}[L] & = \sum_{i=1}^n \mathrm{E}[L_i]\\
\intertext{and}
\mathrm{E}[L] & = \sum_{j=1}^m \mathrm{E}[Y_j].
\end{align}
\end{subequations}
Is there any reasonable way to further decompose the vertical and horizontal expected loss components $\mathrm{E}[L_i]$ and $\mathrm{E}[Y_j]$ respectively into sub-components $E_{k \ell}$ such that
\begin{subequations}
\begin{align}\label{eq:EL_additivity}
\mathrm{E}[L_i] & = \sum_{\ell=1}^m E_{i \ell}\\
\intertext{and}
\mathrm{E}[Y_j] & = \sum_{k=1}^n E_{k j}? \label{eq:EY_additivity}
\end{align}
\end{subequations}
This question is of particular interest in the case where a portfolio of assets may contain names both directly as bonds or loans owed by the names or indirectly by the involvement of the names via CDO tranches. A decomposition of portfolio expected loss as indicated by \eqref{eq:EL_additivity} and \eqref{eq:EY_additivity} can then yield at least a rough measure of a single name's impact on the portfolio loss distribution.
Interestingly enough, Euler's theorem \ref{th:euler} proves very useful also for answering this question.

Observe first that it suffices to find a decomposition of $\mathrm{E}[\min(L,c)]$ where $c$ stands for any credit enhancement level. If then the decomposition procedure is applied consistently for all enhancement levels $c_j$, $j = 1, \ldots, m-1$, property \eqref{eq:EL_additivity} will be satisfied thanks to definition \eqref{eq:Y}. Consider then the problem in the dynamic setting as indicated by \eqref{eq:Lu} and assume that the enhancement level $c$ also depends on the exposure sizes $u = (u_1, \ldots, u_n)$, i.e.\ $c = c(u)$. We are then studying the function $F(u)$ with
\begin{equation}\label{eq:Fu}
    F(u) \ = \ \mathrm{E}[\min(L(u), c(u))].
\end{equation}
From \eqref{eq:Lu} follows that $F$ is homogeneous of degree 1 in the sense of Definition \ref{de:homo} if
$c$ as a function of $u$ is homogeneous of degree 1. Two such choices of $c$ are common in practice:
$c$ as a quantile (cf.\ \eqref{eq:quantile}) of the loss distribution, i.e.\
\begin{subequations}
\begin{align}
    c(u) & = q_\alpha(L(u)) \label{eq:c_quant} \\
\intertext{for some $\alpha \in (0,1)$, or $c$ as some multiple $b>0$ of the portfolio expected loss, i.e.}
c(u) & = b \sum_{i=1}^n u_i\,\mathrm{E}[L_i].\label{eq:c_exp}
\end{align}
\end{subequations}
Homogeneity of degree 1 implies the following solution to the decomposition problem as stated by \eqref{eq:EL_additivity} and \eqref{eq:EY_additivity}.
\begin{proposition}\label{pr:cdo}
Let the credit enhancement levels $c_j = c_j(u)$, $j = 1, \ldots, m-1$ be functions that are homogeneous of degree 1 in the sense of Definition \ref{de:homo}. Let the portfolio loss variable $L(u)$  and the tranche loss variables $Y_j = Y_j(u)$, $j = 1, \ldots, m$ be given by \eqref{eq:Lu} and \eqref{eq:Y} respectively. Assume that the functions $F_j(u) = \mathrm{E}[\min(L(u), c_j(u))]$, $j = 1, \ldots, m-1$ are partially differentiable with respect to all components of $u = (u_1, \ldots, u_n)$. Then the functions $E_{k \ell}(u)$, $k = 1, \ldots, n$, $\ell = 1, \ldots, m$ with
\begin{equation}\label{eq:solution}
    E_{k \ell}(u) \ = \ \left\{
                          \begin{array}{ll}
                           u_k \frac{\partial F_1}{\partial\, u_k}(u), & \hbox{$\ell = 1$;} \\[1ex]
                           u_k \left(\frac{\partial F_\ell}{\partial\, u_k}(u) - \frac{\partial F_{\ell-1}}{\partial\, u_k}(u)\right), &
                                    \hbox{$2 \le \ell \le m-1$;} \\[1ex]
                            u_k \left(\mathrm{E}[L_k] - \frac{\partial F_{m-1}}{\partial\, u_k}(u)\right), & \hbox{$\ell = m$,}
                          \end{array}
                        \right.
\end{equation}
satisfy \eqref{eq:EL_additivity} and \eqref{eq:EY_additivity}.
\end{proposition}
In the following, the quantities $E_{k \ell}(u)$ will be called \emph{tranche loss components}.
Of course, what makes \eqref{eq:solution} attractive as the solution to \eqref{eq:EL_additivity} and \eqref{eq:EY_additivity} is the fact that by inclusion of partial derivatives the resulting contributions to tranche expected losses reflect the potential dynamics from changes in the weights of the underlying names. This fact is illustrated by the following proposition.

\begin{proposition}\label{pr:extreme}
Assume the setting of Proposition \ref{pr:cdo}. If for some $j \in \{1, \ldots, m\}$ there is a portfolio weight vector $u^\ast$ such that $\frac{\mathrm{E}[Y_j(u^\ast)]}{\mathrm{E}[L(u^\ast)]}$ is a local minimum or maximum, then for all $i = 1, \ldots, n$ follows that
\begin{equation*}
    \frac{E_{i j}(u^\ast)}{u_i^\ast\,\mathrm{E}[L_i]} \ = \ \frac{\mathrm{E}[Y_j(u^\ast)]}{\mathrm{E}[L(u^\ast)]}.
\end{equation*}
\end{proposition}
By Proposition \ref{pr:extreme}, in particular, there cannot be a local minimum or maximum of the ratio of tranche expected loss and portfolio expected loss as long as not all of the corresponding ratios of tranche loss components and expected losses of underlying assets are equal.

Note that $F(u) = \mathrm{E}[\min(L(u), c(u))]$ from Proposition \ref{pr:cdo} can be represented as follows:
\begin{subequations}
\begin{align}\label{eq:F_rep}
    F(u) & \ = \ c(u)\,\mathrm{P}[L(u) > c(u)] + \mathrm{E}[L(u)\,\mathbf{1}_{\{L(u) \le c(u)\}}]\\
    & \ = \ c(u)\,\mathrm{P}[L(u) > c(u)] + \mathrm{P}[L(u) \le c(u)]\,\mathrm{E}[L(u)\,|\,L(u) \le c(u)].\label{eq:F_cond}
\end{align}
\end{subequations}
\eqref{eq:F_rep} is suitable for the calculation of the partial derivatives of $F$ in many cases, e.g.\ when $c(u)$ is given by  \eqref{eq:c_exp}. If $c$ is given by \eqref{eq:c_quant}, i.e.\ as a quantile, and the distribution of $L(u)$ is continuous at $q_\alpha(L(u))$ for all $u$, \eqref{eq:F_cond} is more useful.
\begin{remark}\label{rm:F_formula} If the distribution of $L(u)$ is enough smooth with respect to the variable $u$, and $c$ from \eqref{eq:F_rep} is partially differentiable, then the partial derivatives of $F$ can be calculated as follows:
\begin{subequations}
\begin{equation}\label{eq:F_deriv}
    \frac{\partial F}{\partial u_k}(u) \ = \ \frac{\partial\,c}{\partial\, u_k}(u)\, \mathrm{P}[L(u) > c(u)] -
                        c(u)\, \frac{\partial\,\mathrm{P}[L(u) \le c(u)]}{\partial\, u_k}
                    + \frac{\partial\,\mathrm{E}[L(u)\,\mathbf{1}_{\{L(u) \le c(u)\}}]}{\partial\, u_k}.
\end{equation}
In case of $c(u) = q_\alpha(L(u))$ with $q_\alpha(L(u))$ being the quantile function as defined in \eqref{eq:quantile}, we have $\mathrm{P}[L(u) \le c(u)] = \alpha$ for all $u$ if the distribution of $L(u)$ is continuous in $q_\alpha(L(u))$ for all $u$. Taking into account \eqref{eq:VaR_contrib} and \eqref{eq:ES_contrib} then leads to the following simplified version of \eqref{eq:F_deriv}:
\begin{equation}\label{eq:F_quant}
    \frac{\partial\, \mathrm{E}\bigl[\min\bigl(L(u), q_\alpha(L(u))\bigr)\bigr]}{\partial u_k} \ = \ (1-\alpha)\,\mathrm{E}[L_k\,|\,L(u) = q_\alpha(L(u))] +
                    \alpha\,\mathrm{E}[L_k\,|\,L(u) \le q_\alpha(L(u))].
\end{equation}
\end{subequations}
\end{remark}
%
\refstepcounter{figure}
\begin{figure}[tb]
  \begin{center}
  \parbox{14.0cm}{Figure \thefigure:
  \emph{Ratios of tranche loss components and sub-portfolio expected losses (``tranche EL ratios'') as function of weight of first sub-portfolio. Definitions of loss variables as in \eqref{eq:def_Li} and \eqref{eq:def_Lu}. Parameter values as in \eqref{eq:params}. Tranches defined by \eqref{eq:tranches}.}}
\label{fig:1}\\[1ex]
\ifpdf
    \resizebox{\height}{11.0cm}{\includegraphics[width=16.0cm]{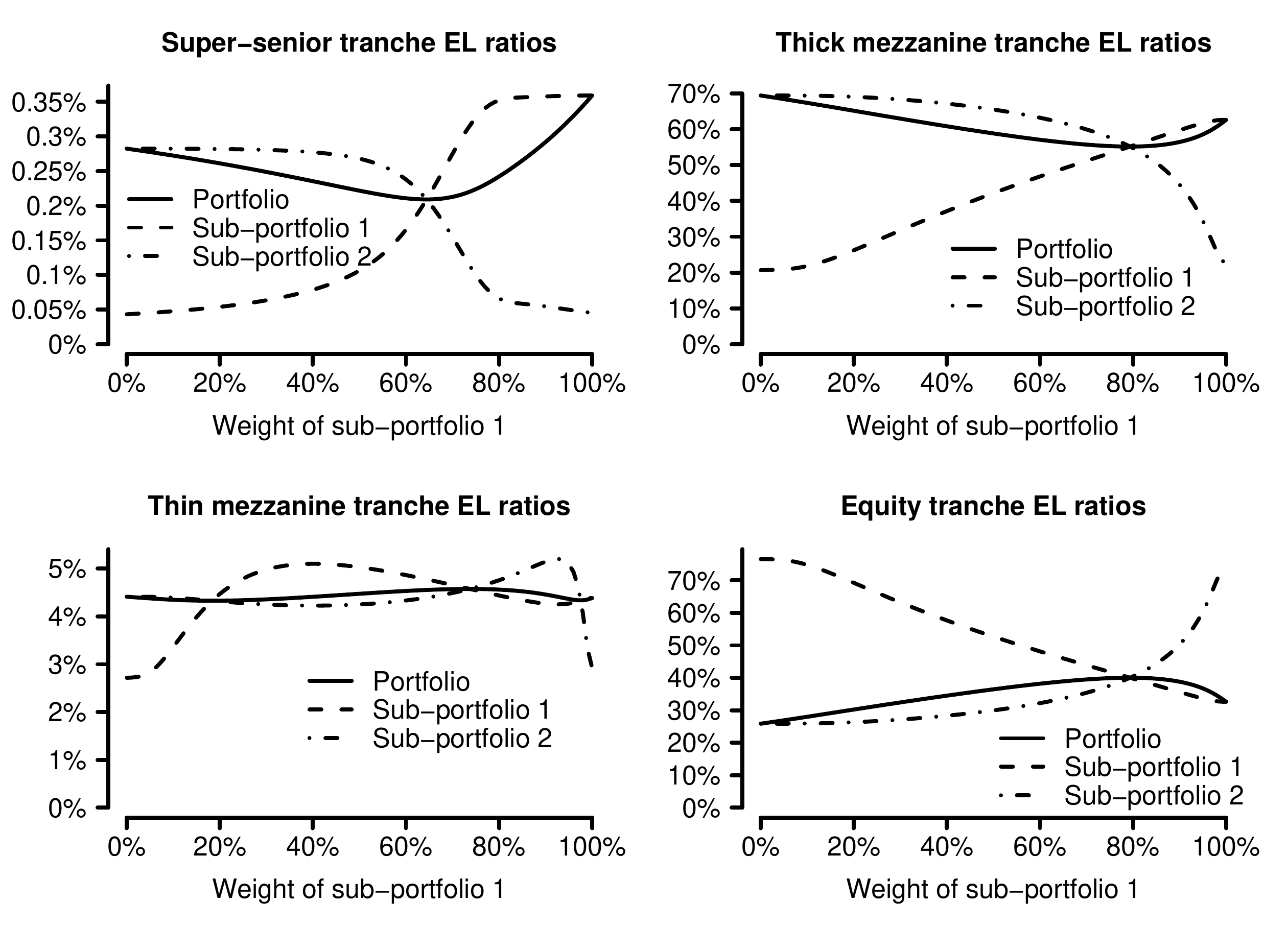}}
\else
\begin{turn}{270}
\resizebox{\height}{12.0cm}{\includegraphics[width=9.0cm]{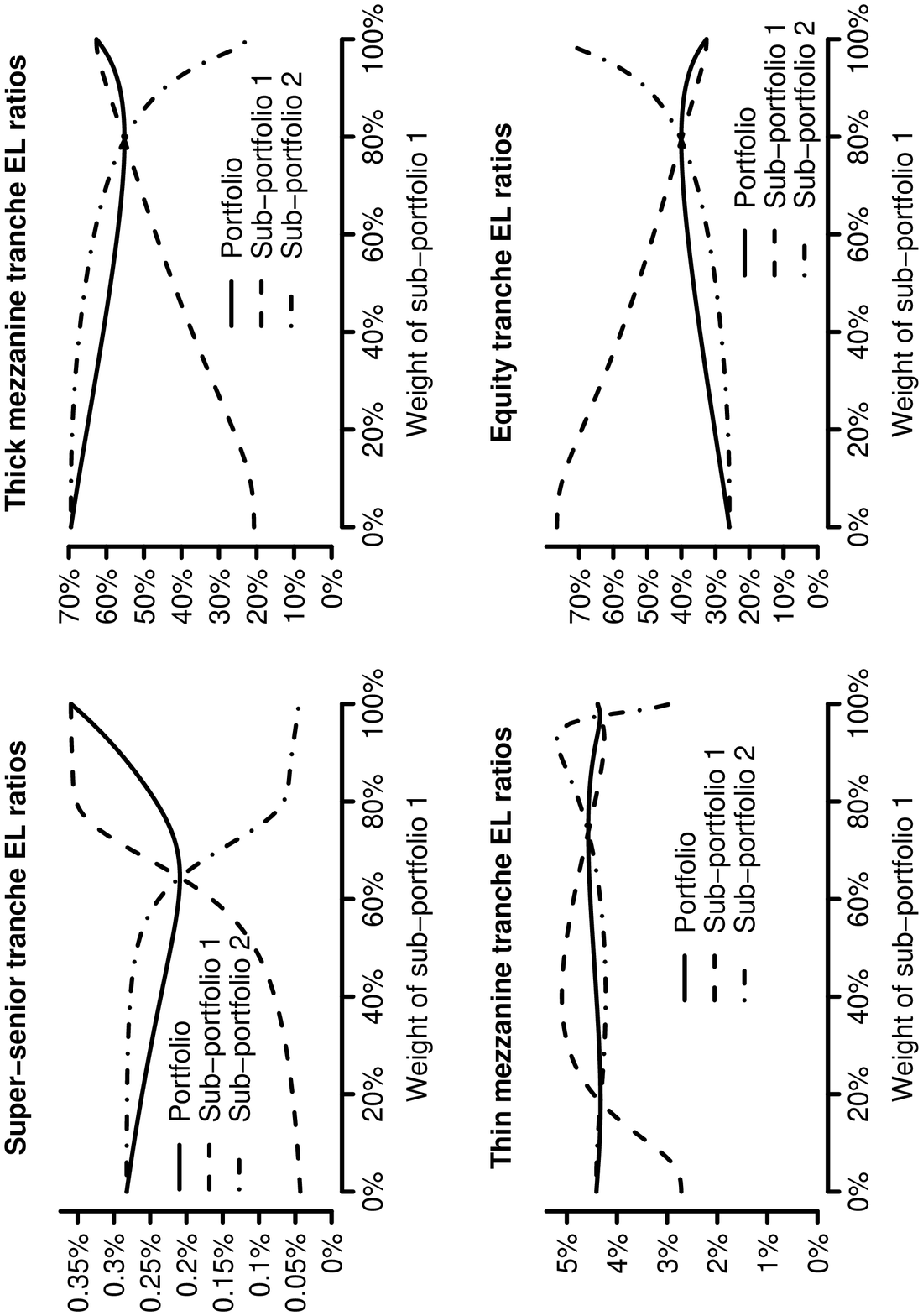}}
\end{turn}
\fi
\end{center}
\end{figure}
We illustrate\footnote{%
The R-scripts for the calculations related to the examples in Sections \ref{se:CDO} and \ref{se:nl_impact} can be down-loaded at \texttt{http://www-m4.ma.tum.de/pers/tasche/}.} the notion of tranche loss component and Proposition \ref{pr:extreme} with an asymptotic two-factor model as in Example~1 of \citet{Tasche2006}. Define $L_i$, $i = 1,2$ as Vasicek-distributed random variables by
\begin{subequations}
\begin{align}
    L_i & = \Phi\left(\frac{t_i + \sqrt{\varrho_i}\,S_i}{\sqrt{1-\varrho_i}}\right),\label{eq:def_Li}
\intertext{with \emph{asset correlations} $\varrho_i \in (0,1)$, \emph{default thresholds} $t_i \in \mathbb{R}$, and \emph{systematic factors} $S_1, S_2$ jointly standard normal such that $\mathrm{corr}[S_1,\,S_2] = \tau\in [0,1]$. Let}
L(u) & = u\,L_1 + (1-u)\,L_2, \quad u \in [0,1].\label{eq:def_Lu}
\end{align}
\end{subequations}
$L(u)$ then can be interpreted as the loss (expressed as fraction of total exposure) in a credit portfolio which consists of two homogeneous sub-portfolios with corresponding loss variables $L_1$ and $L_2$. The parameters in \eqref{eq:def_Li} and \eqref{eq:def_Lu} are chosen as follows:
\begin{gather}
    t_1 = \Phi^{-1}(0.01), \quad \varrho_1 = 0.2 \notag\\
    t_2 = \Phi^{-1}(0.025), \quad \varrho_2 = 0.3 \label{eq:params}\\
    \tau = 0.4.\notag
\end{gather}
We define four tranche loss variables $Y_1(u), Y_2(u), Y_3(u), Y_4(u)$ according to \eqref{eq:Y} by
\begin{equation}\label{eq:tranches}
    \begin{split}
    c_0(u) & = 0,\\
    c_1(u) & = q_{50\%}(L(u)),\\
    c_2(u) & = q_{55\%}(L(u)),\\
    c_3(u) & = q_{99.9\%}(L(u)),\\
    c_4(u) & = 1.
    \end{split}
\end{equation}
For obvious reasons, in the following we call tranche 1 the ``equity tranche'', tranche 2 the ``thin mezzanine tranche'', tranche 3 the ``thick mezzanine tranche'', and tranche 4 the ``super-senior tranche''.
Figure \ref{fig:1} displays for each of the four tranches the ratio of the expected losses of the whole tranche and the whole portfolio as well as the ratios of the tranche loss components and the expected losses of the two sub-portfolios.
Note the different scales of the four charts. Not surprisingly, the EL ratios are quite small for the super-senior and thin mezzanine tranches. Compared to the thicknesses of the tranches, as measured by the differences of the defining confidence levels, the EL ratios of the super-senior and the thick mezzanine tranches are over-proportional. As stated in Proposition \ref{pr:extreme}, in each chart the three curves intersect in the local extremes of the ratios of whole tranche and whole portfolio expected losses. The charts demonstrate that both local minima and maxima may occur and that even both kinds of extremes can be observed in one curve (in the case of the thin mezzanine tranche).

\subsection{Measuring non-linear risk impact}
\label{se:nl_impact}

In Section \ref{sec:theory}, we discussed the problem of how to meaningfully define risk contributions in the case of a \emph{linear} portfolio as given by \eqref{eq:defY} or \eqref{eq:L}. By considering the dynamic modification \eqref{eq:X(u)} of the problem, we argued that only Euler risk contributions \eqref{eq:deriv} can be applied for determining optimal weights for the assets in the portfolio and, at the same time, making sure that the risk contributions add up at portfolio-wide risk. The reasoning for this, however, is heavily based on linearity of the portfolio profit/loss variable with respect to the profit/loss variables of the included assets and on homogeneity of the risk measure. At first glance, we dealt with a non-linear portfolio in Section \ref{se:CDO} when we looked at the tranche loss variables $Y_j$, as defined by \eqref{eq:Y}, which are clearly not linear in the exposure sizes $u_i$ from \eqref{eq:Lu}. But we did not solve in full generality the problem of tranche loss decomposition. Instead we confined ourselves to the case where the credit enhancement levels are defined by functions homogeneous in the exposure sizes. This way, we essentially re-stated the problem in a shape similar to the setting of Section \ref{sec:theory} and accessible by means of Euler's theorem.

In order to analyze non-linear impact of a single factor or a set of factors on portfolio-wide risk, \citet{MartinTasche} suggested to apply Euler allocation to the decomposition of the portfolio-wide profit/loss variable into the expectation of portfolio-wide profit/loss conditional on the factor (or set of factors) and its orthogonal complement. In the rest of the section, we will present this approach in some detail and illustrate it by an example based on the toy portfolio model as specified by \eqref{eq:def_Li} and \eqref{eq:def_Lu}.

\begin{definition}\label{de:impact}
Let $L$ be a loss variable with finite expectation and $\rho$ be a risk measure that is homogeneous of degree 1 in the sense of Definition \ref{de:homo}. Consider a set of risk factors $S = (S_1, \ldots, S_k)$
and the expected value $\mathrm{E}[L\,|\,S]$ of $L$ conditional on $S$. Assume that the Euler contribution $\rho(\mathrm{E}[L\,|\,S]\,|\,L)$ of $\mathrm{E}[L\,|\,S]$ to $\rho(L)$ in the sense of Remark \ref{de:euleralloc} is well-defined and that $\rho(L) > 0$. Then the \emph{risk impact} of the factor set $S$ on $L$ (with respect to $\rho$) is defined as
\begin{equation*}
    \mathrm{RI}_\rho(L\,|\,S) \ = \ \frac{\rho\bigl(\mathrm{E}[L\,|\,S]\,|\,L\bigr)}{\rho(L)}.
\end{equation*}
\end{definition}
\begin{samepage}
\refstepcounter{figure}
\begin{figure}[tb]
\centering
  \parbox{14.0cm}{Figure \thefigure:
  \emph{Risk impacts according to Definition \ref{de:impact} for factors $S_1$ and $S_2$ of the model given by \eqref{eq:def_Li},
  \eqref{eq:def_Lu}, and \eqref{eq:params}, as functions of the weight of the first sub-portfolio. Results for risk impacts based on VaR and ES at 99.9\% confidence level and standard deviation.}}
\label{fig:2}\\[1ex]
\ifpdf
    \resizebox{\height}{10.0cm}{\includegraphics[width=20.0cm]{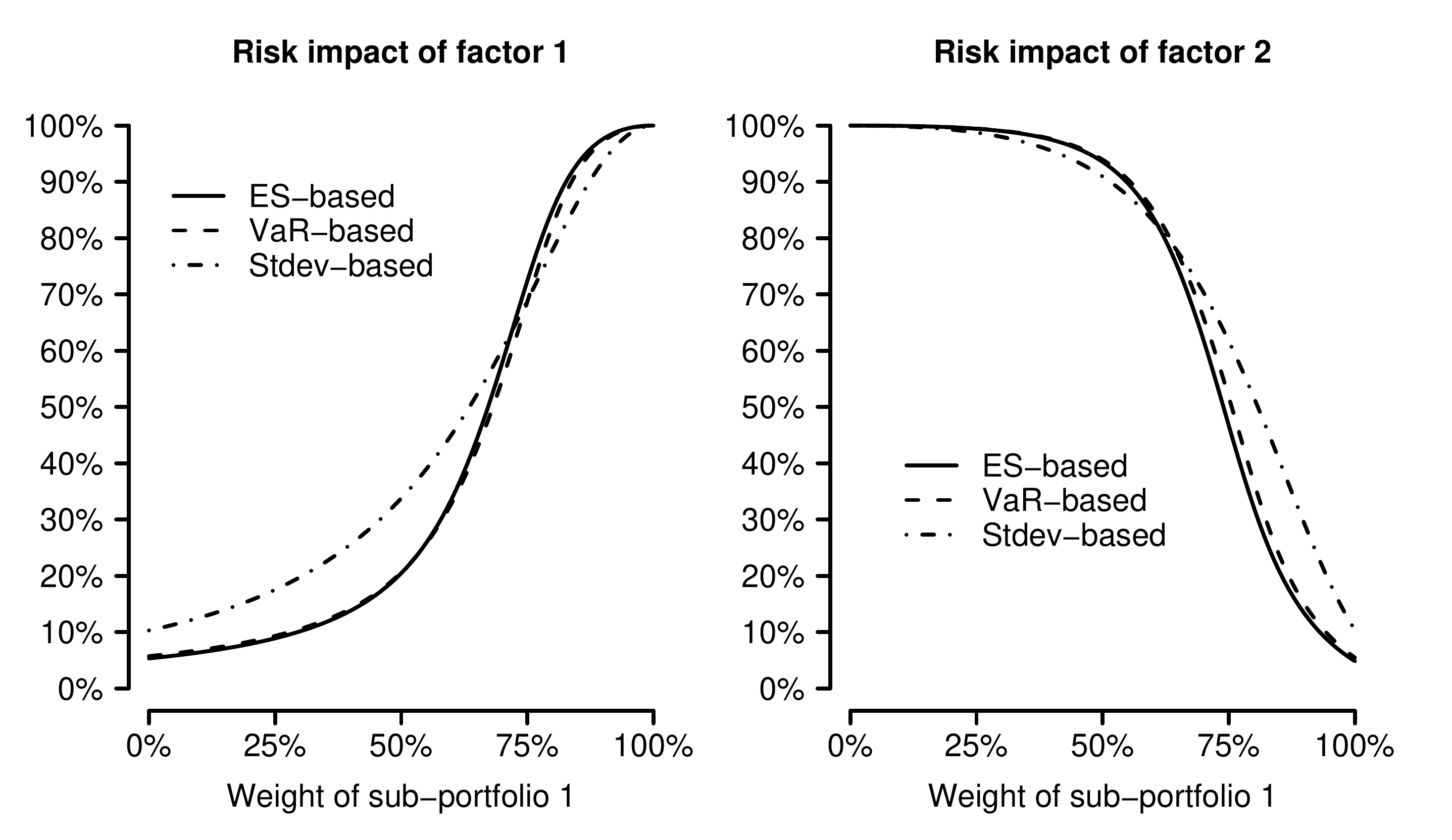}}
\else
\begin{turn}{270}
\resizebox{\height}{12.0cm}{\includegraphics[width=6.0cm]{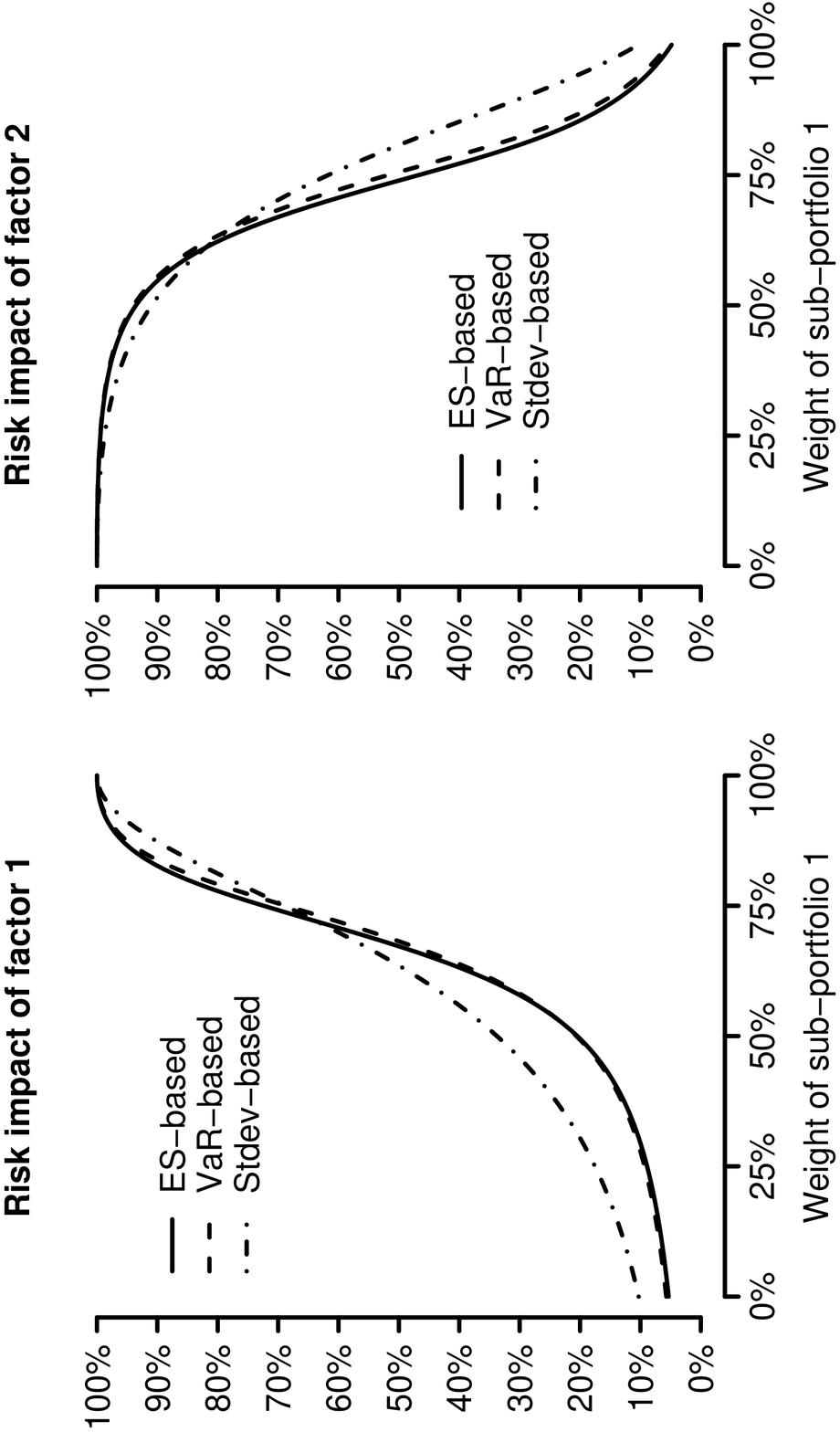}}
\end{turn}
\fi
\end{figure}
\end{samepage}
A key property of conditional expectation is uncorrelatedness of conditional expected value and residual. Applied to the situation of Definition \ref{de:impact}, this means
\begin{subequations}
      \begin{equation}\label{eq:uncorr}
        \mathrm{corr}\bigl[\mathrm{E}[L\,|\,S],\,L-\mathrm{E}[L\,|\,S]\bigr]\ = \ 0.
      \end{equation}
      By the factorization lemma \citep[see, e.g.,][Remark 1.3 (v)]{Stannat2008}, there is a deterministic function $f(s_1, \ldots, s_k)$ such that
      \begin{equation}\label{eq:fact}
        \mathrm{E}[L\,|\,S] \ = \ f(S_1, \ldots, S_k).
      \end{equation}
      Hence the first step in determining the risk impact $\mathrm{RI}_\rho(L\,|\,S)$ essentially means decomposing the portfolio-wide loss variable $L$ into a deterministic function of the factors in $S$ and some uncorrelated residual. Doing so by calculating the expectation of $L$ conditional on $S$ ensures that this decomposition works in an exhaustive manner that also covers non-linear effects.
\end{subequations}

What is the relationship between the  notion of \emph{risk impact} from Definition \ref{de:impact} when applied to the components of a linear portfolio as in \eqref{eq:defY} or \eqref{eq:L} and the notion of \emph{Euler contribution} in the sense of Remark \ref{de:euleralloc}, i.e.\ what is (in the context of \eqref{eq:L}) the connection between $\mathrm{RI}_\rho(L\,|\,L_i)$ and $\rho(L_i\,|\,L)$?

      Let us assume, without loss of generality, that $i=1$. It is instructive to consider the case where $L_1$ and $L_2, \ldots, L_n$ are stochastically independent. Then we obtain
      \begin{subequations}
      \begin{align}
            \mathrm{E}[L\,|\,L_1] & = L_1 - \mathrm{E}[L_1] + \mathrm{E}[L] \\
    \intertext{and hence}
                \rho\bigl(\mathrm{E}[L\,|\,L_1]\,|\,L\bigr) & = \rho\bigl(L_1 - \mathrm{E}[L_1] + \mathrm{E}[L]\,|\,L\bigr).
      \end{align}
      If $\rho$ is a risk measure such that the risk is not changed by adding constants, i.e.
      \begin{equation}\label{eq:trans}
        \rho(X+a)\ = \ \rho(X), \quad a \in \mathbb{R},
      \end{equation}
      then from \eqref{eq:deriv} follows that $\rho\bigl(L_1 - \mathrm{E}[L_1] + \mathrm{E}[L]\,|\,L\bigr) = \rho\bigl(L_1 \,|\,L\bigr)$ and therefore
      \begin{equation}\label{eq:rel}
        \frac{\rho\bigl(L_1 \,|\,L\bigr)}{\rho(L)} \ = \ \mathrm{RI}_\rho(L\,|\,L_i).
      \end{equation}
      Hence, in the case of an independent asset the risk impact coincides with the relative contribution of the asset to portfolio-wide risk.
      In general, without an assumption of independence, one intuitively would expect $\le$ instead of $=$ in \eqref{eq:rel}. Showing this, however, seems non-trivial and presumably would require additional assumptions. We will not enter into this discussion here, but just conclude that, for reasons of compatibility of the definitions of risk contributions and risk impact, it makes sense to impose property \eqref{eq:trans} on the underlying risk measure or even to require that it is a general deviation measure in the sense of \citet{Rockafellaretal2002}.
      \end{subequations}

$\sigma_c$ from Section \ref{se:stddev} is a risk measure with property \eqref{eq:trans}. In this case
  \begin{equation}\label{eq:R2}
    \mathrm{RI}_{\sigma_c}(L\,|\,S) \ = \ \mathrm{RI}_{\sigma}(L\,|\,S)\ = \ \frac{\mathrm{var}\bigl[\mathrm{E}[L\,|\,S]\bigr]}{\mathrm{var}[L]}\
            \in\ [0,1].
  \end{equation}
  According to \eqref{eq:R2}, therefore, risk impact as introduced in Definition \ref{de:impact} is a generalization of the notion of ``coefficient of determination'' or ``$R^2$'' from regression analysis. As regression models can be ranked with respect to their $R^2$ values, with values closer to 100\% indicating a better predictive power, also risk factors related to a portfolio can be ranked with respect to their associated RIs. A higher RI will then reflect a higher impact of the corresponding factor or set of factors that might require stronger attention of the risk managers.

Thinking of risk impact in similar terms as of a coefficient of determination could suggest to define risk impact simply by
  \begin{equation}\label{eq:qRI}
    \mathrm{qRI}_\rho(L|S) \ = \ \frac{\rho(\mathrm{E}[L\,|\,S])}{\rho(L)},
  \end{equation}
  with ``qRI'' standing for \emph{quasi-RI}. This notion, however, is not compatible with Euler contributions in the sense of \eqref{eq:rel}.
  We will show by example that quasi-RI tends to emphasize the impact of a risk factor, when compared to risk impact RI.

Unexpected loss with respect to VaR and ES as defined in \eqref{eq:ULVaR_L} and \eqref{eq:ULES_L} are further examples of risk measures that satisfy property \eqref{eq:trans} and thus define risk impact measures compatible with Euler contributions in the sense of \eqref{eq:rel}. Based on \eqref{eq:VaR_contrib_L} and \eqref{eq:ES_contrib_L}, we obtain the following formulae for risk impact with respect to VaR and ES respectively\footnote{%
      For the sake of keeping the notation concise, we do not explicitly mention ``UL'' in the formulae.}%
      \begin{equation}\label{eq:RI_VaR_ES}
        \begin{split}
            \mathrm{RI}_{\mathrm{VaR}, \alpha}(L\,|\,S) & = \frac{\mathrm{E}\bigl[\mathrm{E}[L\,|\,S]\,|\,L = q_\alpha(L)\bigr] -
                 \mathrm{E}[L]}{q_\alpha(L) - \mathrm{E}[L]}, \\[1ex]
            \mathrm{RI}_{\mathrm{ES}, \alpha}(L\,|\,S) & = \frac{\mathrm{E}\bigl[\mathrm{E}[L\,|\,S]\,|\,L \ge q_\alpha(L)\bigr] -
                 \mathrm{E}[L]}{\mathrm{E}[L\,|\,L \ge q_\alpha(L)] - \mathrm{E}[L]}.
        \end{split}
      \end{equation}
    \citet[][Eq.~(7)]{MartinTasche} proved that $\mathrm{RI}_{\mathrm{ES}, \alpha}(L\,|\,S) \le 1$. To the author's knowledge, no other general bounds to $\mathrm{RI}_{\mathrm{ES}, \alpha}(L\,|\,S)$ or $\mathrm{RI}_{\mathrm{VaR}, \alpha}(L\,|\,S)$ can be derived. Note that
    \begin{equation}\label{eq:ind_case}
        \mathrm{RI}_{\mathrm{ES}, \alpha}(L\,|\,S) = 0,\quad  \mathrm{RI}_{\mathrm{VaR}, \alpha}(L\,|\,S) = 0, \quad \text{and}\quad 
        \mathrm{RI}_{\sigma}(L\,|\,S) = 0
    \end{equation}
    if $L$ and $S$ are independent. 
    
For the two-factor example from Section \ref{se:CDO}, the expectations $\mathrm{E}[L(u)\,|\,S_i]$, $i = 1, 2$  of the portfolio loss variable $L(u)$ given by
\eqref{eq:def_Lu} conditional on the systematic factors $S_1$ and $S_2$ can easily be determined. A short calculation shows that
\begin{equation}\label{eq:cond_exp}
   \begin{split}
   \mathrm{E}[L(u)\,|\,S_1] & = u\,\Phi\left(\frac{t_1 + \sqrt{\varrho_1}\,S_1}{\sqrt{1-\varrho_1}}\right) +
        (1-u)\,\Phi\left(\frac{t_2 + \sqrt{\varrho_2}\,\tau\,S_1}{\sqrt{1-\varrho_2\,\tau^2}}\right),\\
   \mathrm{E}[L(u)\,|\,S_2] & = u\,\Phi\left(\frac{t_1 + \sqrt{\varrho_1}\,\tau\,S_2}{\sqrt{1-\varrho_1\,\tau^2}}\right) +
        (1-u)\,\Phi\left(\frac{t_2 + \sqrt{\varrho_2}\,S_2}{\sqrt{1-\varrho_2}}\right).
   \end{split} 
\end{equation}
Based on these representations of $\mathrm{E}[L(u)\,|\,S_i]$, $i = 1, 2$, formulae \eqref{eq:R2}, \eqref{eq:RI_VaR_ES}, and \eqref{eq:qRI} (for ES) can be numerically evaluated. Figure \ref{fig:2} displays some results for VaR-based, ES-based, and standard deviation-based risk impacts (VaR and ES at 99.9\% confidence level). For the specific example under consideration, the VaR-based and ES-based curves can hardly be distinguished. The standard deviation-based curves are somehow flatter than the quantile-based curves, indicating a slightly softer transition from low to high impact of the factors. For both factors, nonetheless, all curves demonstrate a relatively abrupt transition from low impact at low weight of the sub-portfolio directly depending on the factor to high impact at high weight of that sub-portfolio. Not surprisingly, due to the specific choice \eqref{eq:params} of the model parameters which renders the second sub-portfolio more risky, the spectrum of low impact weights is much wider for the first sub-portfolio.
Note that, as a consequence of the positive correlation of the two factors, both factors have positive risk impact even when the weight of their corresponding sub-portfolios is naught.

Figure \ref{fig:3} again displays the $\mathrm{RI}_{\mathrm{ES}, \alpha}$-curves from Figure \ref{fig:2}, but this time together with the corresponding $\mathrm{qRI}_{\mathrm{ES}, \alpha}$ curves of the two factors. For both factors, the quasi-RI curves suggest significantly higher risk impacts than the RI curves. As noticed before, however, the ranking of the factors implied by the risk impact values is more interesting than the values themselves. According to Figure \ref{fig:3}, the implied rankings are essentially identical because both the quasi-RI curves as well as the RI-curves seem to intersect at the same sub-portfolio 1 weight of roughly 71\%. Actually, closer inspection shows that the quasi-RI curves intersect at a weight not greater than 70.61\% while the RI curves criss-cross at a weight of not less than 70.70\%. Similar behaviour of the curves appears also to obtain with other values of the parameters in \eqref{eq:params}. It hence would be interesting to see whether calculations for more realistic portfolios yield similar results. For the time being, the conclusion might be that the choice between risk impacts and quasi-RIs should mainly depend on how much numerical effort is required for the different approaches.
\begin{samepage}
\refstepcounter{figure}
\begin{figure}[tb]
\centering
  \parbox{14.0cm}{Figure \thefigure:
  \emph{Risk impacts according to Definition \ref{de:impact} and quasi-RIs according to \eqref{eq:qRI} for factors $S_1$ and $S_2$ of the model given by \eqref{eq:def_Li},
  \eqref{eq:def_Lu}, and \eqref{eq:params}, as functions of the weight of the first sub-portfolio. Results for ES at 99.9\% confidence level.}}
\label{fig:3}\\[1ex]
\ifpdf
    \resizebox{\height}{4.0in}{\includegraphics[width=4.0in]{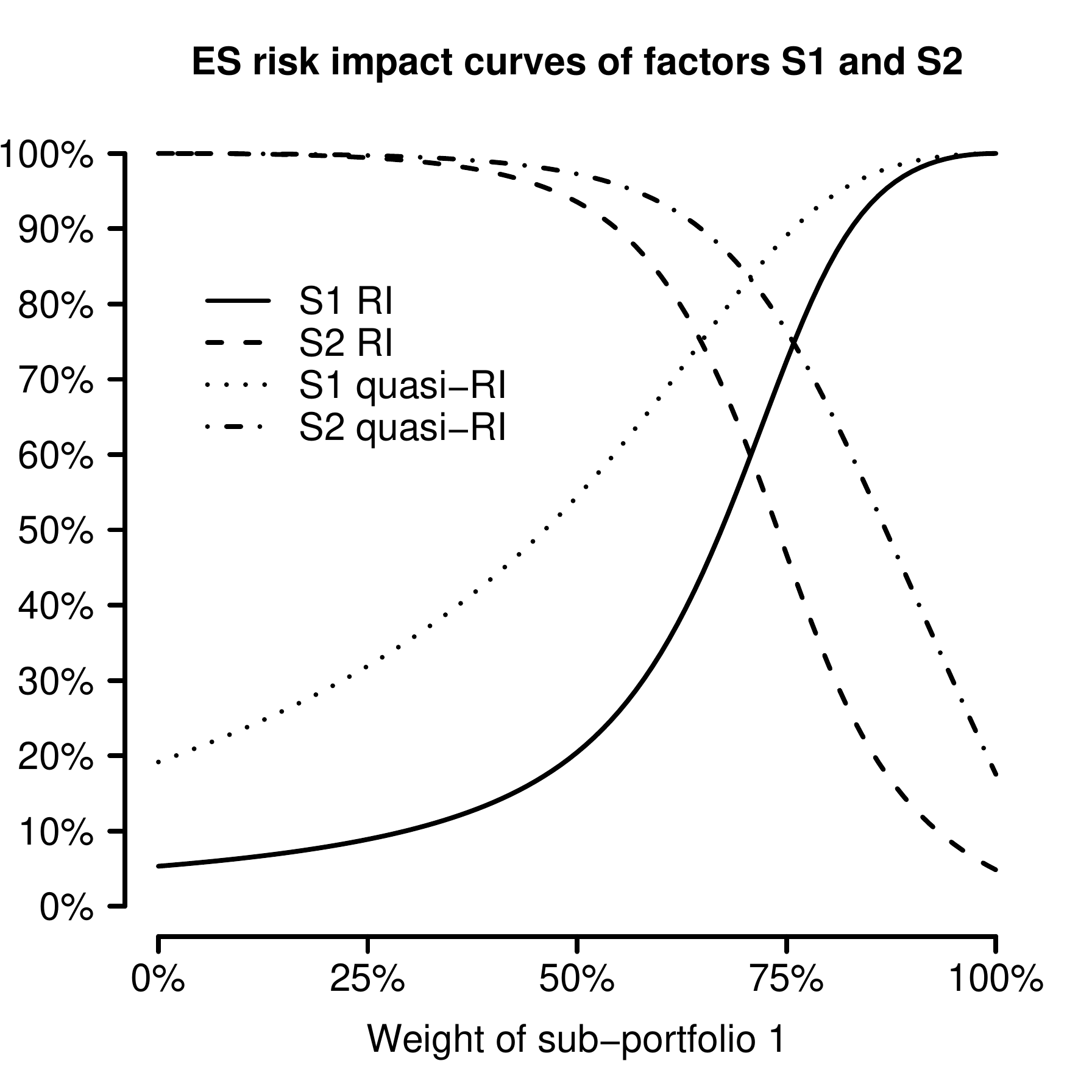}}
\else
\begin{turn}{270}
\resizebox{\height}{9.0cm}{\includegraphics[width=9.0cm]{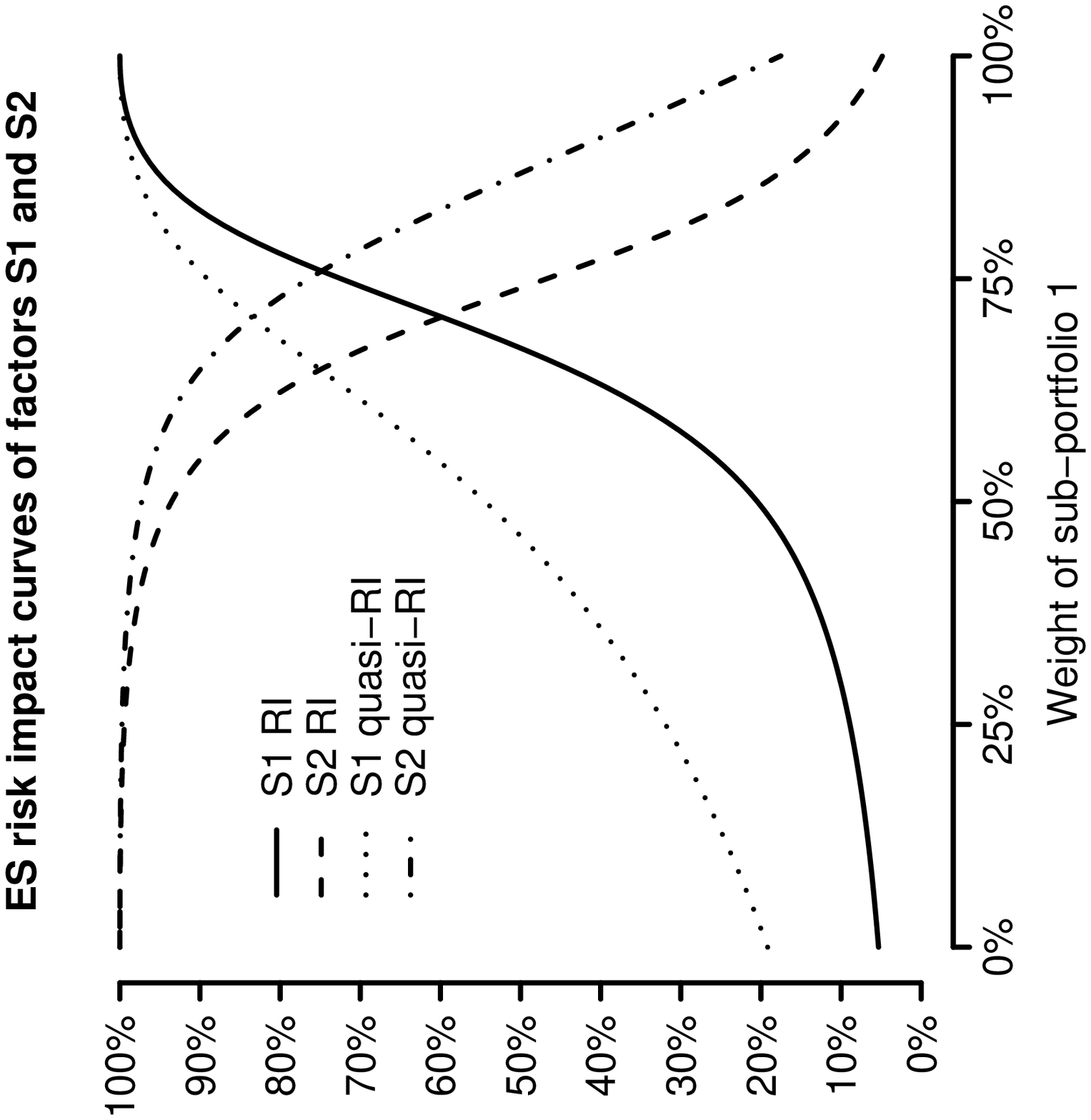}}
\end{turn}
\fi
\end{figure}
\end{samepage}

\section{Summary}
\label{sec:concl}

Among the many methodologies that financial institutions apply for their internal capital allocation process the so-called \emph{Euler allocation} is especially appealing for its economic foundation. There are a lot of papers that provide partial information on the Euler allocation but a comprehensive overview seems to be missing so far. In this chapter, we have tried to fill this gap to some extent. In particular, besides emphasizing the economic meaning of Euler allocation, we have discussed its potential application to the detection of risk concentrations. We have moreover demonstrated that there is a natural relationship between Euler contributions to VaR and kernel estimators for conditional expectations.

As two further, somehow different applications of Euler's theorem, but still in the context of portfolio risk, on the one hand we have introduced a new approach to the decomposition of CDO tranche expected losses into loss components that can be associated with the underlying names. On the other hand, we have revisited a concept introduced by \citet{MartinTasche} for measuring the impact of systematic or other risk factors on portfolio-wide risk.



\begin{thebibliography}{38}
\providecommand{\natexlab}[1]{#1}
\providecommand{\url}[1]{\texttt{#1}}
\expandafter\ifx\csname urlstyle\endcsname\relax
  \providecommand{\doi}[1]{doi: #1}\else
  \providecommand{\doi}{doi: \begingroup \urlstyle{rm}\Url}\fi

\bibitem[Acerbi and Tasche(2002)]{Acerbi&Tasche}
C.~Acerbi and D.~Tasche.
\newblock On the coherence of expected shortfall.
\newblock \emph{Journal of Banking \& Finance}, 26\penalty0 (7):\penalty0
  1487--1503, 2002.

\bibitem[Artzner et~al.(1999)Artzner, Delbaen, Eber, and Heath]{ADEH99}
P.~Artzner, F.~Delbaen, J.-M. Eber, and D.~Heath.
\newblock Coherent measures of risk.
\newblock \emph{Mathematical Finance}, 9\penalty0 (3):\penalty0 203--228, 1999.

\bibitem[BCBS(2006)]{BaselAccord}
BCBS.
\newblock \emph{{I}nternational {C}onvergence of {C}apital {M}easurement and
  {C}apital {S}tandards. {A} {R}evised {F}ramework, {C}omprehensive {V}ersion}.
\newblock {B}asel {C}ommittee of {B}anking {S}upervision, June 2006.

\bibitem[Bluhm(2003)]{Bluhm2003}
C.~Bluhm.
\newblock C{DO} {M}odeling: {T}echniques, {E}xamples and {A}pplications.
\newblock Working paper, HypoVereinsbank, 2003.

\bibitem[Denault(2001)]{Denault01}
M.~Denault.
\newblock Coherent allocation of risk capital.
\newblock \emph{Journal of Risk}, 4\penalty0 (1):\penalty0 1--34, 2001.

\bibitem[Dev(2004)]{Dev2004}
A.~Dev, editor.
\newblock \emph{Economic Capital: A Practitioner Guide}, 2004. {R}isk {B}ooks.

\bibitem[Dhaene et~al.(2006)Dhaene, Vanduffel, Tang, Goovaerts, Kaas, and
  Vyncke]{Dhaeneetal}
J.\ Dhaene, S.\ Vanduffel, Q.\ Tang, M.\ Goovaerts, R.\ Kaas, and D.~Vyncke.
\newblock Risk measures and comonotonicity: a review.
\newblock \emph{Stochastic Models}, 22\penalty0 (4):\penalty0 573--606, 2006.

\bibitem[Epperlein and Smillie(2006)]{EpperleinSmillie}
E.\ Epperlein and A.~Smillie.
\newblock Cracking {VAR} with kernels.
\newblock \emph{RISK}, 19\penalty0 (8):\penalty0 70--74, August 2006.

\bibitem[Fischer(2003)]{Fischer}
T.~Fischer.
\newblock Risk capital allocation by coherent risk measures based on one-sided
  moments.
\newblock \emph{Insurance: Mathematics and Economics}, 32\penalty0
  (1):\penalty0 135--146, 2003.

\bibitem[Garcia~Cespedes et~al.(2006)Garcia~Cespedes, de~Juan~Herrero, Kreinin,
  and Rosen]{Garcia04}
J.~C.\ Garcia~Cespedes, J.~A.\ de~Juan~Herrero, A.\ Kreinin, and D.~Rosen.
\newblock A simple multifactor ``factor adjustment'' for the treatment of
  credit capital diversification.
\newblock \emph{Journal of Credit Risk}, 2\penalty0 (3):\penalty0 57--85, 2006.

\bibitem[Glasserman(2005)]{Glasserman2005}
P.~Glasserman.
\newblock Measuring {M}arginal {R}isk {C}ontributions in {C}redit {P}ortfolios.
\newblock \emph{Journal of Computational Finance}, 9:\penalty0 1--41, 2005.

\bibitem[Gouri\'{e}roux et~al.(2000)Gouri\'{e}roux, Laurent, and
  Scaillet]{GL00}
C.~Gouri\'{e}roux, J.~P. Laurent, and O.~Scaillet.
\newblock Sensitivity analysis of values at risk.
\newblock \emph{Journal of Empirical Finance}, 7:\penalty0 225--245, 2000.

\bibitem[Gr\"undl and Schmeiser(2007)]{Grundl2007}
H.~Gr\"undl and H.~Schmeiser.
\newblock Capital allocation for insurance companies: What good is it?
\newblock \emph{Journal of Risk \& Insurance}, 74\penalty0 (2):\penalty0
  301--317, 2007.

\bibitem[Kalkbrener(2005)]{Kalkbrener05}
M.~Kalkbrener.
\newblock An axiomatic approach to capital allocation.
\newblock \emph{Mathematical Finance}, 15\penalty0 (3):\penalty0 425--437,
  2005.

\bibitem[Kalkbrener et~al.(2004)Kalkbrener, Lotter, and
  Overbeck]{Kalkbreneretal04}
M.\ Kalkbrener, H.\ Lotter, and L.~Overbeck.
\newblock Sensible and efficient allocation for credit portfolios.
\newblock \emph{{RISK}}, 17:\penalty0 S19--S24, January 2004.

\bibitem[Koyluoglu and Stoker(2002)]{Koyluoglu&Stoker2002}
U.\ Koyluoglu and J.~Stoker.
\newblock Honour your contribution.
\newblock \emph{RISK}, 15\penalty0 (4):\penalty0 90--94, April 2002.

\bibitem[Litterman(1996)]{Litterman96}
R.~Litterman.
\newblock Hot spots$^\mathrm{TM}$ and hedges.
\newblock \emph{The Journal of Portfolio Management}, 22:\penalty0 52--75,
  1996.

\bibitem[Martin and Tasche(2007)]{MartinTasche}
R.\ Martin and D.~Tasche.
\newblock Shortfall: a tail of two parts.
\newblock \emph{{RISK}}, 20\penalty0 (2):\penalty0 84--89, February 2007.

\bibitem[Mc{N}eil et~al.(2005)Mc{N}eil, Frey, and Embrechts]{McNeil05}
A.\ Mc{N}eil, R.\ Frey, and P.~Embrechts.
\newblock \emph{Quantitative {R}isk {M}anagement}.
\newblock Princeton University Press, 2005.

\bibitem[Memmel and Wehn(2006)]{MemmelWehn}
C.\ Memmel and C.~Wehn.
\newblock The supervisor's portfolio: the market price risk of {G}erman banks
  from 2001 to 2004 -- {A}nalysis and models for risk aggregation.
\newblock \emph{Journal of Banking Regulation}, 7:\penalty0 309--324, 2006.

\bibitem[Merino and Nyfeler(2004)]{Merino04}
S.\ Merino and M.~A. Nyfeler.
\newblock Applying importance sampling for estimating coherent credit risk
  contributions.
\newblock \emph{Quantitative Finance}, 4:\penalty0 199--207, 2004.

\bibitem[Myers and Read(2001)]{MyersRead}
S.~C.\ Myers and J.~A. Read.
\newblock Capital allocation for insurance companies.
\newblock \emph{The Journal of Risk and Insurance}, 68:\penalty0 545--580,
  2001.

\bibitem[Pagan and Ullah(1999)]{Pagan99}
A.\ Pagan and A.~Ullah.
\newblock \emph{Nonparametric econometrics}.
\newblock Cambridge University Press, 1999.

\bibitem[Patrik et~al.(1999)Patrik, Bernegger, and R\"uegg]{Patriketal}
G.\ Patrik, S.\ Bernegger, and M.\ R\"uegg.
\newblock The use of risk adjusted capital to support business decision making.
\newblock \emph{Casualty Actuarial Society Forum}, 1999.

\bibitem[Rockafellar and Uryasev(2002)]{Rockafellar&Uryasev}
R.~T. Rockafellar and S.~Uryasev.
\newblock Conditional {V}alue-at-{R}isk for general loss distributions.
\newblock \emph{Journal of Banking \& Finance}, 26\penalty0 (7):\penalty0
  1443--1471, 2002.

\bibitem[Rockafellar et~al.(2002)Rockafellar, Uryasev, and
  Zabarankin]{Rockafellaretal2002}
R.~T. Rockafellar, S.~Uryasev, and M.~Zabarankin.
\newblock Deviation {M}easures in {R}isk {A}nalysis and {O}ptimization.
\newblock Research report 2002-7, I{SE} {D}ept., University of Florida, 2002.

\bibitem[Schwaiger(2006)]{Schwaiger2006}
W.~Schwaiger.
\newblock Controlling risikobasierter {E}rfolge in {U}niversalbanken.
\newblock Working paper, Institut f\"ur Managementwissenschaften, Technische
  Universit\"at Wien, 2006.

\bibitem[Stannat(2008)]{Stannat2008}
W.~Stannat.
\newblock Probability theory.
\newblock Lecture notes, Technische Universit\"at Darmstadt, 2008.
\newblock Fifth part -- corrected version.

\bibitem[Tasche(1999)]{Tasche1999}
D.~Tasche.
\newblock Risk contributions and performance measurement.
\newblock Working paper, Technische Universit\"at M\"unchen, 1999.

\bibitem[Tasche(2002)]{Tasche2002}
D.~Tasche.
\newblock Expected {S}hortfall and {B}eyond.
\newblock \emph{Journal of Banking and Finance}, 26\penalty0 (7):\penalty0
  1519--1533, 2002.

\bibitem[Tasche(2004{\natexlab{a}})]{Tasche2004a}
D.~Tasche.
\newblock Allocating portfolio economic capital to sub-portfolios.
\newblock In A.~Dev, editor, \emph{Economic Capital: A Practitioner Guide},
  pages 275--302. {R}isk {B}ooks, 2004{\natexlab{a}}.

\bibitem[Tasche(2004{\natexlab{b}})]{Tasche2004}
D.~Tasche.
\newblock Capital {A}llocation with {C}redit{R}isk$^+$.
\newblock In V.~M.\ Gundlach and F.~B. Lehrbass, editors, \emph{CreditRisk$^+$
  in the Banking Industry}, pages 25--44. Springer, 2004{\natexlab{b}}.

\bibitem[Tasche(2006)]{Tasche2006}
D.~Tasche.
\newblock Measuring sectoral diversification in an asymptotic multifactor
  framework.
\newblock \emph{Journal of Credit Risk}, 2\penalty0 (3):\penalty0 33--55, 2006.

\bibitem[Tasche(2007)]{Tasche2007}
D.~Tasche.
\newblock Capital allocation for credit portfolios with kernel estimators.
\newblock Working paper, 2007.

\bibitem[Tasche and Tibiletti(2004)]{TascheTibiletti}
D.\ Tasche and L.~Tibiletti.
\newblock Approximations for the {V}alue-at-{R}isk approach to risk-return
  analysis.
\newblock \emph{The ICFAI Journal of Financial Risk Management}, 1\penalty0
  (4):\penalty0 44--61, 2004.

\bibitem[Turlach(1993)]{Turlach}
B.~Turlach.
\newblock Bandwidth {S}election in {K}ernel {D}ensity {E}stimation: {A}
  {R}eview.
\newblock Discussion paper 9317, Institut de Statistique, Universit\'e
  Catholique de Louvain, 1993.

\bibitem[Urban et~al.(2004)Urban, Dittrich, Kl\"uppelberg, and
  St\"olting]{Urbanetal2004}
M.\ Urban, J.\ Dittrich, C.\ Kl\"uppelberg, and R.~St\"olting.
\newblock Allocation of risk capital to insurance portfolios.
\newblock \emph{Bl\"atter der DGVFM}, 26:\penalty0 389--406, 2004.

\bibitem[Yamai and Yoshiba(2002)]{Yamai02}
Y.\ Yamai and T.~Yoshiba.
\newblock Comparative {A}nalyses of {E}xpected {S}hortfall and {V}a{R}: their
  estimation error, decomposition, and optimization.
\newblock Monetary and economic studies 20(1), Bank of Japan, 2002.

\end{thebibliography}

\appendix

\section{Euler's theorem on homogeneous functions}
\label{sec:eulertheorem}

In this chapter, the focus is on \emph{homogeneous} risk measures and functions.
\begin{definition}\label{de:homo}
A risk measure $\rho$ in the sense of \eqref{eq:ec} is called \emph{homogeneous of degree $\tau$}
if for all $h > 0$ the following equation obtains:
\begin{equation*}
    \rho(h\,X)\ =\ h^\tau\,\rho(X).
\end{equation*}
A function $f: U \subset \mathbb{R}^n \to \mathbb{R}$ is called \emph{homogeneous of degree $\tau$} if for
all $h > 0$ and $u\in U$ with $h\,u \in U$ the following equation holds:
\begin{equation*}
    f(h\,u)\ =\ h^\tau\,f(u).
\end{equation*}
\end{definition}
Note that the function $f_\rho$ corresponding by \eqref{eq:fct_risk} to the risk measure $\rho$ is
homogeneous of degree $\tau$ if $\rho$ is homogeneous of degree $\tau$.

In the case of continuously differentiable functions, homogeneous functions can be described by Euler's theorem.
\begin{theorem}[Euler's theorem on homogeneous functions]\label{th:euler}
Let $U\subset \mathbb{R}^n$ be an open set and $f: U \to \mathbb{R}$ be a continuously differentiable function. Then
$f$ is homogeneous of degree $\tau$ if and only if it satisfies the following equation:
\begin{equation*}
    \tau\,f(u)\ = \sum_{i=1}^n u_i\,\frac{\partial\,f(u)}{\partial\,u_i}, \qquad u = (u_1,\ldots, u_n) \in U,\ h > 0.
\end{equation*}
\end{theorem}
\begin{remark}
It is easy to show that $\frac{\partial\,f}{\partial\,u_i}$ is homogeneous of degree $\tau -1$ if $f$ is homogeneous
of degree $\tau$. From this observation follows that if $f$ is homogeneous of degree 1 and continuously differentiable for $u = 0$
then $f$ is a linear function (i.e.\ with constant partial derivatives). Often, therefore, the homogeneous functions relevant
for risk management are not differentiable in $u=0$.
\end{remark}

Functions $f: U\subset \mathbb{R}^n\to \mathbb{R}$ that are homogeneous of degree 1 are \emph{convex}, i.e.
\begin{subequations}
\begin{equation}\label{eq:convex}
f(\eta\,u + (1-\eta)\,v) \ \le \ \eta\,f(u) + (1-\eta)\,f(v), \qquad u, v\in U,\ \eta \in [0,1],
\end{equation}
if and only if they are \emph{sub-additive}, i.e.
\begin{equation}\label{eq:fctsubadd}
    f(u + v) \ \le \ f(u) + f(v), \qquad u,v \in U.
\end{equation}
\end{subequations}
Theorem \ref{th:euler} then implies the following useful characterization of sub-additivity for continuously differentiable functions
that are homogeneous of degree 1.

\begin{corollary}\label{co:sub}
Let $U\subset \mathbb{R}^n$ be an open set and $f: U \to \mathbb{R}$ be a continuously differentiable function that
is homogeneous of degree 1. Then $f$ is sub-additive (and convex) if and only if the following inequality holds:
\begin{equation*}
    \sum_{i=1}^n u_i\,\frac{\partial\,f(u+v)}{\partial\,u_i}\ \le \ f(u), \qquad u, u+v \in U.
\end{equation*}
\end{corollary}
See Proposition
2.5 of \citet{Tasche2002} for a proof of Corollary \ref{co:sub}.

\end{document}